\documentclass[12pt]{article}
\usepackage{latexsym}
\usepackage[pdftex]{graphicx}
\usepackage{amsfonts}
\usepackage{amsmath, amsthm, amssymb}
\usepackage{fullpage}
\usepackage{subcaption}
\usepackage{setspace}
\usepackage{multirow}
\usepackage{longtable}
\usepackage{authblk}
\usepackage{natbib}
\bibpunct{(}{)}{;}{a}{}{,}
\usepackage{dcolumn}
\newcolumntype{.}{D{.}{.}{-1}}
\newcolumntype{d}[1]{D{.}{.}{#1}}
\usepackage[top=1in, left=.9in, right=.9in,
bottom=1in]{geometry}
\usepackage{bm}
\usepackage[compact]{titlesec}
\usepackage{booktabs}
\usepackage{tikz, pgfplots}
\usetikzlibrary{positioning}
\usetikzlibrary{shapes.arrows}
\usetikzlibrary{decorations.pathreplacing,calligraphy}
\usepackage{color}
\RequirePackage[colorlinks,citecolor=blue,urlcolor=blue]{hyperref}
\usepackage{amsthm}
\newtheorem{prop}{Proposition}
\newtheorem{assumption}{Assumption}
\newtheorem{theorem}{Theorem}

\def\var{\mbox{Var}}
\def\bmD{\bm{D}}

\def\bmY{\bm{Y}}
\def\bmO{\bm{O}}
\def\calM{{\cal M}}
\def\bmV{\bm{V}}
\def\bmv{\bm{v}}
\def\bpsi{\bm{\psi}}
\def\bmeta{\bm{\eta}}
\def\D{\bm{D}}
\def\Y{\bm{Y}}

\def\X{\bm{X}}
\def\x{\bm{x}}

\expandafter\def\expandafter\normalsize\expandafter{\normalsize\setlength\abovedisplayskip{0pt}}
\expandafter\def\expandafter\normalsize\expandafter{\normalsize\setlength\belowdisplayskip{0pt}}
\expandafter\def\expandafter\normalsize\expandafter{\normalsize\setlength\abovedisplayshortskip{0pt}}
\expandafter\def\expandafter\normalsize\expandafter{\normalsize\setlength\abovedisplayshortskip{0pt}}

\begin{document}
\pagestyle{plain}

\title{Instrumented Difference-in-Differences}
\author{Ting Ye\footnote{Department of Statistics, Wharton School, University of Pennsylvania. }, Ashkan Ertefaie\footnote{Department of Biostatistics and Computational Biology, University of Rochester.}, James Flory\footnote{Department
		of Subspecialty Medicine, Memorial Sloan Kettering Cancer Center.}, Sean Hennessy\footnote{Department of Biostatistics, Epidemiology, and Informatics, Perelman School of Medicine,
		University of Pennsylvania}, and Dylan S. Small\footnote{Address for Correspondence: Dylan S. Small, Department of Statistics, Wharton School, University of Pennsylvania, 3730 Walnut Street, Philadelphia, PA 19104, U.S.A. (E-mail: {\tt dsmall@wharton.upenn.edu}).\\ This work was supported by grant R01AG064589 from the National Institutes of Health.  The work is not peer-reviewed. } }

\date{}

\maketitle
\thispagestyle{empty}

\abstract{
Unmeasured confounding is a key threat to reliable causal inference based on observational studies. Motivated from two powerful natural experiment devices,  the instrumental variables and difference-in-differences, we propose a new method called instrumented difference-in-differences that explicitly leverages exogenous randomness in an exposure trend  to estimate the average and conditional average treatment effect in the presence of unmeasured confounding. %Specifically, we use an instrumental variable for DID, a variable that (i) is associated with trend in exposure; (ii) is independent of the potential exposures, 
%potential trends in outcome  and individual treatment effect; and (iii) has no direct effect on the trend in outcome and does not modify the individual treatment effect.  
We develop the identification assumptions using the potential outcomes framework. We propose a Wald estimator and a class of multiply robust and efficient semiparametric estimators, with provable consistency and asymptotic normality. In addition, we extend the  instrumented difference-in-differences to a two-sample design to facilitate investigations of delayed treatment effect and provide a measure  of weak identification.
We demonstrate our results in simulated and real datasets.

}

\newcommand{\n}{\noindent}

\clearpage

\section{Introduction}
\label{sec: intro}
Unmeasured confounding is a key threat to reliable causal inference based on observational studies \citep{Lawlor:2004aa, Rutter:2007aa}. A popular approach to handle unmeasured confounding  is 
the instrumental variable (IV) method, which requires an IV that satisfies three core assumptions \citep{Angrist:1996aa, Baiocchi:2014aa, Hernan-Robins}: 
 (i) (relevance) it is associated with the exposure; (ii) (independence) it is independent of any unmeasured confounder of the exposure-outcome relationship; (iii) (exclusion restriction) it has no direct effect on the outcome.  By extracting exogenous variation in the exposure that is independent of the unmeasured confounder, IVs can be used to estimate the causal effect. 

%
%For example, \cite{Newman:2012aa} used hospital's preference for phototherapy when treating newborns with hyperbilirubinemia  as an IV to study the effect of phototherapy, and pointed out that hospital's preference may have a direct effect on the outcome as hospitals with a higher preference of phototherapy also have a greater use of infant formula. 

Meanwhile, the increasing availability of large longitudinal datasets such as administrative claims and electronic health records has created new opportunities to expand study designs to take advantage of the longitudinal structure. One method that is widely used  in economics and other social sciences is difference-in-differences (DID) \citep{Card:1994aa, Angrist:2008aa}. The method of DID is based on a comparison of the trends in outcome for two exposure 
groups, where one group consists of  individuals who switch from being unexposed to exposed and the other group consists of individuals who are never exposed. Under the parallel trends assumption, which says that the outcomes in the two exposure groups  evolve in the same way over time in the absence of the exposure, DID is able to  remove time-invariant bias from the unmeasured confounder.  However, because the setup and assumptions of DID are motivated from applications in social sciences, its applicability is limited in biomedical sciences. For example, in social sciences it is relatively common for  a new policy to be applied to one region of the country but not another, creating a circumstance in which key assumptions such as parallel trends are likely to hold and facilitating a DID design. In assignment of pharmacologic or other treatments in health care, such clear natural, exogenous sources of cleavage between exposed and unexposed groups are rare, making it more difficult to identify situations in which all assumptions of DID will be met. 

In this article, we connect these two powerful natural experiment devices  (referred to as the standard IV and standard DID) and propose a new method called instrumented DID  to estimate 
the causal effect of the exposure in the presence of  unmeasured confounding. Unlike the standard DID, the instrumented DID exploits a \emph{haphazard} encouragement targeted at a subpopulation towards faster uptake of the exposure or a surrogate of such encouragement, which we call \emph{IV for DID}. Then any observed nonparallel trends in outcome between the encouraged and unencouraged groups  provides
evidence for causation, as long as  their trends in outcome are parallel if all individuals were counterfactually not  exposed. 
 A prototypical example of instrumented DID is a longitudinal randomized experiment, where after a baseline period, some individuals are randomly selected to be encouraged to take the treatment regardless of their treatment history. If the encouragement is effective, the exposure rate would increase more for the encouraged group than the unencouraged group. If additionally the encouragement has no direct effect on the trend in outcome, then 
 any nonparallel trends in outcome must be due to the nonparallel trends in exposure. Therefore, through exploiting  haphazard encouragement that affects the exposure trend, the instrumented DID is able to extract some variation in the exposure trend that is independent of the unmeasured confounder and relax some of the most disputable assumptions of the standard IV and standard DID method, particularly the exclusion restriction for the standard IV method and the parallel trends for the standard DID method; see Section \ref{sec: iv for trend} for more discussion.

 Reasoning similar to the instrumented DID has been applied informally in prior studies. A prominent example is the differential trends in smoking prevalence for men and women as a consequence of targeted tobacco advertising to women, which were associated with disproportional trends for men and women in lung cancer mortality \citep{Burbank:1972aa, Meigs:1977aa, Patel:2004aa}. Specifically, because of marketing efforts designed to introduce specific women's brands of cigarettes such as Virginia Slims in 1967, there was a considerable increase in smoking initiation by young women, which lasted through the mid-1970s \citep{pierce1995historical}. Thirty years later, the lung cancer mortality rates for women 55 or older had increased to almost four times the 1970 rate, whereas rates among men had no such dramatic change \citep{Bailar:1997aa}.  In Section \ref{sec: data}, we will analyze this example using the proposed method.

The rest of this paper is organized as follows. In Section \ref{sec: iv for trend}, we introduce notation and our setup, and  establish the identification assumptions for the instrumented DID using a potential outcomes framework. In Section \ref{sec: semi}, we develop a Wald estimator and  a class of semiparametric efficient estimators, and derive their asymptotic properties. In Section \ref{sec: two-sample}, we extend the instrumented DID to a two-sample design. In Section \ref{sec: weak}, we provide a measure of weak identification. Results from simulation studies and a real data application are in Sections \ref{sec: simu} and \ref{sec: data}, respectively. The paper concludes with a discussion in Section \ref{sec: discussions}. We implement the proposed method in the   \textsf{R} package \textsf{idid}, available at {\tt https://github.com/tye27/idid}.

\vspace{-8mm}

\section{Instrumented DID: Setup, Potential Outcomes, Causal Effect, Identification}
\label{sec: iv for trend}

Suppose that we observe an independent and identically distributed (i.i.d.) sample $ (\bmO_1,\dots, $  $\bmO_n) $ with $\bmO=( T,  Z, \X, D, Y)$, where $ T $ is a binary time indicator which equals $t$ if an observation  is from time $t$, $ Z $ is a  binary IV for DID observed at the baseline, $\X$ is a vector of baseline covariates, $ D $ is a binary exposure variable, $ Y $ is some real-valued outcome of interest. For defining causal effects, we use the potential outcomes framework \citep{Neyman:1923a, Rubin:1974}.  Define $D_t^{(z)}$ as the potential exposure that would be observed at time $t$ if  $Z$ were externally set to $z$, define $Y_t^{(d)}$ as the potential outcome that would be observed at time $t$ if $ D $ were externally set to $ d $ and $ Z $ had the same value it actually had. The full data vector for each individual  is $(Z, \X, D_t^{(z)}, Y^{(d)}_t, t=0,1, z=0,1, d=0,1)$. Moreover, let  $ Y^{(d)} := Y_T^{(d)}$ be the potential outcome that would be observed if $ D $ were externally set to $ d $, and $ T $ and $ Z $ were set to the values that naturally occur.  Our goal is to make inferences about the average treatment effect $ 	\beta_0= E(Y^{(1)}-Y^{(0)} ) $ and the conditional average treatment effect $ 	\beta_0(\bmv)= E(Y^{(1)}-Y^{(0)} | \bmV=\bmv), $
where $ \bmV $ is a pre-specified subset of $ \X $, representing the effect modifiers of interest; for example, setting $ \bmV $ to be an empty set gives the unconditional average treatment effect $ \beta_0 $.  Note that the separation of $ V $ and $ X $ separates the need to adjust for possible confounding and the specification of effect modifiers of interest, which provides great flexibility and allows researchers to define the estimand of interest a priori. Throughout the article, we consider the treatment effect on the additive scale.

We make the following identification assumptions for using the instrumented DID.
\begin{assumption}\label{assum: 1}
	(a) (consistency) $D=D_T^{(Z)}$ and $Y=Y_T^{(D)}$.\\
	(b) (positivity) $ 0<P(T=t, Z=z| \X)<1 $ for $ t=0,1, z=0,1 $ with probability 1. \\
	(c) (random sampling)	$T\perp (D_t^{(z)}, Y^{(d)}_t, t=0,1, z=0,1, d=0,1)| Z, \X$. 
\end{assumption}
Assumption \ref{assum: 1}(a) states that the observed exposure is $ D=D_t^{(z)} $  if and only if $ Z=z $ and $ T=t $, and the observed outcome is $ Y= Y_t^{(d)} $ if and only if $ D=d$ and  $  T=t $. Implicit in this assumption is that an individual's observed outcome is not affected by others' exposure level or this individual's exposure level at the other time point; this is known as  the Stable Unit Treatment Value Assumption \citep{Rubin:1978aa, Rubin:1990aa}. Assumption \ref{assum: 1}(b) postulates that there is a positive probability of receiving each $ (t, z )$ combination within each level of $ \X $, or equivalently, the support of $ \X $ is the same for each level of $ (T,Z) $. Assumption \ref{assum: 1}(c) is often assumed for repeated cross-sectional datasets and says that for each level of $ (Z, \X)$, the collected data at every time point is a random sample from the underlying population; see, for example, Section 3.2.1 of  \cite{abadie2005semiparametric} that makes a similar assumption. 
%For every individual, the absence of data at some time points can be conceptualized as a missing data problem and Assumption \ref{assum: 1}(c) is an analogue of the missing at random assumption \citep{Rubin:1974}. 

\begin{assumption}[Instrumented DID] \label{assump: iv for trend} With probability 1, \hfill
	
	\noindent
	(a) (Trend relevance) $E(D_1^{(1)} - D_0^{(1)} |  Z=1, \X)\neq E(D_1^{(0)} - D_0^{(0)} |  Z=0, \X) $.\\
	(b) (Independence \& exclusion restriction) $Z\perp (D_{t}^{(0)}, D_{t}^{(1)},\! Y_1^{(0)}-Y_0^{(0)}, Y_t^{(1)}-Y_t^{(0)}, t=0,1)| \X$.	\\
	(c) (No unmeasured common effect modifier)	$Cov(D_t^{(1)}-D_t^{(0)},  Y_t^{(1)}-Y_t^{(0)}|\X)=0$ for $  t=0,1$. \\
	(d)  (Stable treatment effect over time)   $ E( Y_1^{(1)}-Y_1^{(0)}| \X)=E(Y_0^{(1)}-Y_0^{(0)}| \X) $. 
\end{assumption}
Assumptions \ref{assump: iv for trend}(a)-(b)  formalize the core assumptions that an IV for DID needs to satisfy and are parallel to the core assumptions for the standard IV introduced in Section \ref{sec: intro} \citep{Angrist:1996aa, tan2006regression,small2007sensitivity, Wang:2018aa}.

Assumption \ref{assump: iv for trend}(a) says that the IV for DID $ Z $, as an encouragement that disproportionately acts on only a subpopulation,  affects the trend in exposure. For example, $ Z $ can be a random encouragement for some subjects in a longitudinal experiment, an advertisement campaign targeted at a certain geographic region or subpopulation, or a change in reimbursement policies for a certain insurance plan.  Under Assumption \ref{assum: 1}, Assumption \ref{assump: iv for trend}(a) is equivalent to $ E(D| T=1, Z=1, \X)-E(D| T=0, Z=1, \X) \neq E(D| T=1, Z=0, \X)-E(D| T=0, Z=0, \X)$ with probability 1, thus is checkable from observed data. 

Assumption \ref{assump: iv for trend}(b) is an integration of the independence and exclusion restriction assumption. To see this, we adopt a more elaborated definition of the potential outcomes and  define $ Y_t^{(dz)} $ as the potential outcome that would be observed at time $ t $ if $ D $ were externally set to $ d $ and $ Z $ to $ z $, then Assumption  \ref{assump: iv for trend}(b) is implied by (independence) $Z\perp (D_{t}^{(0)}, D_{t}^{(1)}, Y_1^{(0z)}-Y_0^{(0z)}, Y_t^{(1z)}-Y_t^{(0z)}, t=0,1, z=0,1)| \X$ and (exclusion restriction) $ Y_t^{(11)}  - Y_t^{(01)} | X \sim_d Y_t^{(10)}  - Y_t^{(00)}| X  $ and $ Y_1^{(01)}-Y_0^{(01)}| X\sim_dY_1^{(00)}-Y_0^{(00)}|X$, where $ \sim_d $ means having the same distribution; see \cite{tan2006regression} for a parallel statement for the standard IV and \cite{Hernan:2006aa} for connections and differences between different definitions of the standard IV. Hence, Assumption \ref{assump: iv for trend}(b) essentially states that $ Z $ is unconfounded, has no direct effect on the trend in outcome, and does not modify the average treatment effect. Here, we see the main advantage of using $ Z $ as an IV for DID compared to as a standard IV: $ Z $ as an IV for DID is allowed to have a direct effect on the outcome, as long as it has no direct effect on the trend in outcome and does not modify the average treatment effect.  For example, Newman et al. (2012) considered using a hospital's preference for phototherapy when treating newborns with hyperbilirubinemia as a standard IV to study the effect of phototherapy but found evidence that hospitals that use more phototherapy also have greater use of infant formula, which is thought to be an effective treatment for hyperbilirubinemia. Hence, the hospital's preference is a potentially invalid standard IV as it can have a direct effect on the outcome through the use of infant formula. However, it may still qualify as an IV for DID if the use of phototherapy evolves differently between the high and low preference hospitals over time, but the use of infant formula in the two groups of hospitals does not change over time.  These features imply that variables like hospital's preference may be more likely to be an IV for DID, compared to being a standard IV.

Assumption \ref{assump: iv for trend}(c) is developed in  \cite{Cui:2020aa} and a slightly stronger version is proposed earlier in \cite{Wang:2018aa}. To better understand this assumption, suppose in this paragraph only 
the existence of an unmeasured confounder $ U_t $ such that $ (D_t^{(1)}, D_t^{(0)}) \perp ( Y_t^{(1)}, Y_t^{(0)})\mid (U_t, \X)$. Then Assumption \ref{assump: iv for trend}(c) holds if either (i) there is no additive $ U_t $-$ Z$ interaction in $ E(D_t| Z, U_t, \X) $: $ E(D_t| Z=1, U_t, \X)- E(D_t| Z=0, U_t, \X) = E(D_t| Z=1, \X)- E(D_t| Z=0, \X) $; or (ii) there is no additive $ U_t$-$d $ interaction in $ E(Y^{(d)}| U_t, \X) $: $ E(Y^{(1)} - Y^{(0)}| U_t, \X) =  E(Y^{(1)} - Y^{(0)}| \X) $. 

Assumption \ref{assump: iv for trend}(d) requires that the average treatment effect does not change over time. This is a strong assumption but may be plausible in many applications when the study period only spans a short period of time. In our application in Section \ref{sec: data}, we conduct a sensitivity analysis to gauge the  sensitivity of the study conclusion to violation of this assumption. 

Two additional remarks on Assumption \ref{assump: iv for trend} are in order. First, an attractive feature of Assumptions  \ref{assump: iv for trend}(c)-(d) is that they are guaranteed to be true under the sharp null hypothesis of no treatment effect for all individuals. This means that the instrumented DID method can be used for testing the sharp null hypothesis under Assumptions   \ref{assump: iv for trend}(a)-(b). Second, according to Assumption  \ref{assump: iv for trend}, the IV for DID $ Z $ is assumed to be causal for the exposure as it is required to be independent of $ (D_{t}^{(1)}, D_t^{(0)}) $ conditional on $ \X $. In the supplementary materials (Section S3), we present another version of Assumption \ref{assump: iv for trend} which does not require $ Z $ to be causal, i.e., $ Z $ is allowed to be correlated with a cause that affects the trend in exposure, and is more suitable for our application in Section \ref{sec: data} in which we use gender as the IV for DID for its correlation with the encouragement from targeted tobacco advertising.

For $ C\in \{Y, D\} $, let $ \mu_C(t, z,\X)= E(C|T=t, Z=z, \X)$, $ \delta_C(\X)= \mu_C(1,1, \X)-\mu_C(0,1,\X)-\mu_C(1,0,\X)+\mu_C(0,0,\X)$, and  let $ \mu_C(t, z) $ and $ \delta_C $ denote their counterparts without observed covariates. The next proposition indicates that the (conditional) average treatment effect can be identified under the above assumptions.  

\begin{prop}  \label{prop: iv for trend, Wald}
	If  Assumptions  \ref{assum: 1}-\ref{assump: iv for trend} hold, then
	\begin{align}
		\frac{\delta_Y(\X)}{\delta_D(\X)} = \beta_0 (\X) \text{   and   }  E\bigg[\frac{\delta_Y(\X)}{\delta_D(\X)}\mid  \bmV=\bmv\bigg] = \beta_0(v).   \label{eq: iv for trend, Wald}
	\end{align} 
\end{prop}

Now we contrast the instrumented DID with the standard DID.  The standard DID compares the trends in outcome between two exposure groups, where every individual in one group switches from being unexposed to exposed between two time points, and every individual in the other group is never exposed. Its key assumption, called the parallel trends, says that the potential outcomes for the two exposure groups would evolve parallelly  in the absence of the exposure, which is violated if there exists time-varying unmeasured confounding in the exposure-outcome relationship.  In contrast,  the instrumented DID explicitly probes the relationship between the trend in outcome and the trend in exposure using an exogenous variable $ Z $  which often results in partial compliance with exposure within groups defined by levels of $ Z $. 
%The key identification assumption in DID, known as the parallel trends assumption, which says the outcomes for the treated and the control groups would have exhibited the same trends over time if untreated, corresponds to the  unconfounded and exclusion restriction assumption in  Assumptions  \ref{assump: iv for trend}(b)-(c). Under the  special design in DID, Assumptions  \ref{assump: iv for trend}(a) and \ref{assump: iv for trend}(d) hold automatically, and therefore, $ \beta $ in (\ref{eq: iv for trend, Wald}) identifies the average treatment effect for the treated at the post-treatment time period.  
Therefore, compared with the standard DID, the instrumented DID is robust to time-varying unmeasured confounding in the exposure-outcome relationship by making use of an exogenous variable $ Z $ that is not subject to this  time-varying unmeasured confounding. %This allows one to make causal inference using the IV for DID in the presence of time-varying. %Furthermore, in the same way as the DID strengthens the direct outcome comparison, we have discussed in details after Assumption \ref{assump: iv for trend} that the IV for trend method strengthens the standard IV method. 

%\begin{table}
%	\caption{Three common study designs and their IV analogues. The standard IV produces an instrumented treated-vs.-control outcome comparison, while as a special instance of the standard IV, the calendar time IV produces an instrumented pre-vs.-post outcome comparison. The IV for trend produces an instrumented Difference-in-Differences.}\label{tb: summary}
%	\centering	
%	\begin{tabular}{ccc} \\ [-2ex] \toprule
%		Common   study design                           &  & IV analogue                         \\[0.5ex] 		\cline{1-1} \cline{3-3} \\[-1.5ex]
%		treated-vs.-control outcome   comparison        &  &  \multirow{2}{*}{ standard IV }     \\
%		(﻿cohort design)                                &  &                                   \\
%%		&  &                                   \\[-2ex] 
%%		pre-vs.-post outcome   comparison               &  & \multirow{2}{*}{calendar time IV} \\
%%		(﻿interrupted time series   design)             &  &                                   \\
%		&  &                                     \\[-2ex]
%		difference-in-differences                       &  & \multirow{2}{*}{IV for trend} \\ 
%		(﻿comparative interrupted time   series design) &  &         \\[0.5ex] \bottomrule
%	\end{tabular}
%\end{table}

%in the sense that DID requires the treated and control groups to satisfy the parallel trends assumption while IV for trend only requires that to hold for the encouragaed and unencouraged groups.

We remark that when there are no observed covariates, $ \delta_Y/\delta_D $ has been derived in alternative ways in econometrics under different assumptions. It is the same as the standard IV Wald ratio after first differencing the exposure and outcome when each individual is observed at both time points \citep[Chapter 15.8]{wooldridge2010econometric},  as motivated from the linear structural equation models. Importantly, Proposition \ref{prop: iv for trend, Wald} 
provides a justification of this approach using the potential outcomes framework without any modeling assumption. It is also the same as the Wald ratio in the fuzzy DID method for identification of a local average treatment effect under the assumption that individuals can switch treatment in only one direction within each treatment group \citep{Chaisemartin:2017aa}, as motivated from social science applications (e.g., \cite{duflo2001schooling}). Compared with this derivation, our proposed instrumented DID is less stringent in terms of the direction in which each individual can switch treatment, thus is better suited for applications using healthcare data where individuals can switch treatment in any direction. In addition, we complement the proposed instrumented DID with a novel semiparametric estimation and inference method in Section \ref{sec: semi}, two-sample design in Section \ref{sec: two-sample}, and measure of weak identification in Section \ref{sec: weak}. 

Finally, we note that Assumption \ref{assump: iv for trend}(c) can be  replaced by the monotonicity assumption, that is $ D_t^{(1)}\geq D_t^{(0)} $ for $ t=0,1 $ with probability 1, under which we show that $\delta_Y(X)/\delta_D(X)$ in (\ref{eq: iv for trend, Wald}) identifies a complier average treatment effect; see Section S2.2 of the supplementary materials for details.

\section{Estimation and Inference}
\label{sec: semi}

\subsection{Wald estimator}
When there are no observed covariates and based on Proposition \ref{prop: iv for trend, Wald}, we can simply replace the conditional expectations in (\ref{eq: iv for trend, Wald}) with their sample analogues and obtain the Wald estimator 
\begin{align}
	\hat{\beta}= \frac{\hat{\mu}_Y(1,1)-\hat{\mu}_Y(0,1)-\hat{\mu}_Y(1,0)+\hat{\mu}_Y(0,0)}{ \hat{\mu}_D(1,1)-\hat{\mu}_D(0,1)-\hat{\mu}_D(1,0)+\hat{\mu}_D(0,0)}= \frac{\hat{\delta}_Y}{\hat{\delta}_D}, \label{eq: Wald estimator}
\end{align}
where $ \hat{\mu}_C(t,z)= \sum_{i=1}^n C_i I(T_i=t, Z_i=z) /\sum_{i=1}^n I(T_i=t, Z_i=z) $, $ \hat{\delta}_C= \hat{\mu}_C(1,1)-\hat{\mu}_C(0,1)-\hat{\mu}_C(1,0)+\hat{\mu}_C(0,0)$, for $ C\in\{Y,D\} $. In Theorem S1 of the supplementary materials, we prove consistency and asymptotic normality of $ \hat\beta $ and give a consistent variance estimator.

\subsection{Semiparametric theory and multiply robust estimators}

Consider the case with a baseline observed covariate vector $ \X$. Suppose that we have a parametric model for $\beta_0(\bmv)$,  written as $\beta(\bmv; \bpsi)$ for some finite-dimensional parameter $\bpsi$. We do not assume this model is necessarily correct, but instead treat it as a working model. Then we use the weighted least squares projection given by 
\begin{align}
	\bpsi_0=\arg\min_{\bpsi} E\left[ w(\bmV) \left\{ \beta_0(\bmV) - \beta(\bmV; \bpsi)\right\}^2\right], \label{eqn: obj}
\end{align}
where $w(\bmv)$ is a user-specified weight function, which can be tailored if there is subject matter knowledge for emphasizing specific parts of the support of $\bmV$; otherwise, we can set $ w(\bmv) =1$. By definition,  $ \beta(\bmV; \bpsi_0) $ is the best least squares approximation to the conditional average treatment effect $ \beta_0(\bmV) $. For example, when  effect modification is not of interest, we can specify $ \beta(\bmv; \bpsi)= \bpsi $ and $ \beta_0(\bmV) $ is projected onto a constant $ \bpsi_0 $, which can be interpreted as the average treatment effect; if we want to estimate a linear approximation of the conditional average treatment effect, we can specify $ \beta(\bmv; \bpsi)= \bmv^T \bpsi	 $, with $ \bmV $ including the intercept. This approach is also adopted in \cite{Abadie:2003aa, Ogburn:2015aa} and \cite{Kennedy:2019aa}.

%We estimate the parameter $ \bpsi_0 $ using the semiparametric approach \citep{Bickel:1993book, vanderVaart:2000book}. 
%First, semiparametric estimators allow for double or multiple robustness, in the sense that  estimators are consistent provided that a subset of the nuisance parameters is correctly specified. Second, semiparametric doubly or multiply robust approaches enable fast root $ n $ convergence rate even when the nuisance parameters are estimated at slower rates. This appealing feature has sparked recent research on using flexible machine learning methods to estimate the nuisance parameters \citep{Chernozhukov:2018aa}. Third, when the nuisance parameters are estimated at fast enough rates, the resulting estimator reaches the semiparametric efficiency bound and is locally efficient.
The next theorem derives the efficient influence function for $ \bpsi $ \citep{Bickel:1993book, vanderVaart:2000book}.

\begin{theorem} \label{theo}
	If Assumptions \ref{assum: 1}-\ref{assump: iv for trend} hold, and $ \partial \beta(\bmv; \bpsi) /\partial \bpsi$ exists and is continuous. 	Under a nonparametric model, \!the efficient influence function for $\bpsi$ is proportional to 
	\begin{align}
		& \varphi(\bmO; \bpsi, \bmeta)=q(\bmV; \bpsi) \left(\frac{\delta_Y(\X)}{ \delta_D(\X)}-\beta(\bmV; \bpsi) \right.	 \label{eq: IF}\\
		& \left.+\frac{(2Z-1) (2T-1)}{\pi(T, Z, \X) \delta_D(\X)}\bigg[  Y-\mu_Y(T,Z,\X)-\frac{\delta_Y(\X)}{ \delta_D(\X)} \{  D-\mu_D(T,Z,\X)\}  \bigg]\right), \nonumber
	\end{align}
	where  $ \mu_Y, \mu_D, \delta_Y, \delta_D $ are defined in Proposition \ref{prop: iv for trend, Wald}, $ \pi(t, z, \x)= P(T=t, Z=z|\X=\x)$, 	$ \bmeta=(\mu_D, \mu_Y, \pi) $ denotes the vector of nuisance parameters,  and $ q(\bmv; \bpsi) = w(\bmv)\partial \beta(\bmv; \bpsi) /\partial \bpsi$.
\end{theorem}
Notice that the efficient influence function gives an estimator $ \hat{\bpsi} $ defined as a solution to  
\begin{align}
	\sum_{i=1}^n \varphi (\bmO_i;
	 \bpsi, \hat{\bmeta})=0, \label{eq: semi estimator}
\end{align}
where $ \hat{\bmeta} =(\hat \mu_D, \hat \mu_Y,\hat \pi)  $ is a vector of estimated nuisance parameters. As an important special case, the estimator $ \hat{\bpsi} $ has an explicit form when the working model is specified to be linear (including the case when $ \beta(\bmV;  \bpsi) = \bpsi$, with $ \bmV =1$). Specifically,
\begin{align}
	\hat{\bpsi}&= \left\{ \sum_{i=1}^n w(\bmV_i) \bmV_i \bmV_i^T\right\}^{-1} \left\{ \sum_{i=1}^{n} w(\bmV_i) \bmV_i  \bigg( \frac{\hat\delta_Y(\X_i)}{ \hat\delta_D(\X_i)} \right. \nonumber\\
	&  \qquad\left. +\frac{(2Z_i-1) (2T_i-1)}{\hat\pi(T_i, Z_i, \X_i) \hat\delta_D(\X_i)}\bigg[  Y_i-\hat\mu_Y(T_i,Z_i,\X_i)-\frac{\hat\delta_Y(\X_i)}{ \hat\delta_D(\X_i)} \left\{ D_i-\hat\mu_D(T_i,Z_i,\X_i)\right\}  \bigg]  \bigg)\right\}.\nonumber
\end{align}

Next we derive the asymptotic properties of $ \hat{\bpsi} $ defined by (\ref{eq: semi estimator}). Consider  three models:
\begin{itemize}
	\item [$\calM_1$]: models for $\pi(t, z, \x), \mu_D(t, z, {\x})$ are correct.
	\item  [$\calM_2$]: models for $\pi(t, z, {\x}), \delta_Y(\x)/\delta_D(\x) $ are correct.
	\item [$\calM_3$]: models for $\mu_Y(t, z,  \x), \mu_D(t,z, {\x})$ are correct.
\end{itemize}
It is proved in the supplementary materials that our estimator $ \hat{\bpsi} $ is multiply robust, in the sense that the estimator is consistent as long as either one of the three models  ($\calM_1, \calM_2, \calM_3$) holds.  %We remark that the multiple robustness property is conceptually different from the double robustness property that has been widely discussed, see, for example,  \cite{Scharfstein:1999aa, Bang:2005aa, Kang:2007aa, Tan:2010aa, Ogburn:2015aa} and \cite{Kennedy:2019aa}. Usually, the doubly robust estimators are consistent when either one of the two components of the likelihood is correctly specified, while our multiply robust estimator is consistent when any one of the three model combinations is correctly specified, where the model combinations may have overlaps. %For example, both $ \calM_1 $ and $ \calM_3 $ require that the exposure model $ \mu_D(t, z, {\x}) $ is correct, and the model for $ \beta( \x) $ in $ \calM_2 $ is also related to $ \mu_D(t, z, {\x}) $. 
%Nonetheless, the multiple robustness property is important because the multiply robust estimator $ \hat \bpsi $ can achieve faster convergence rate even when the nuisance parameters are estimated at slower rates.  
More examples of multiply robust estimators in other settings can be found in \cite{Vansteelandt:2008aa, Wang:2018aa} and \cite{Shi:2020aa}.

Let $ \xrightarrow{p} $ denote convergence in probability, $ \|\bpsi\| = (\bpsi^T \bpsi )^{1/2}$ denote the Euclidean norm,  $ \|f\|_2 = \{\int f^2(\bm o) dP(\bm o) \}^{1/2}$ denote the $ L_2(P) $ norm, where $ P $ denotes the distribution of $ \bmO $, and $ \bmeta_0= (  \mu_{D0}, \mu_{Y0}, \pi_0) $ denote the true values of the nuisance parameters.
\begin{assumption} \label{assump: semi}
	(a) $(\hat{\bpsi},\hat{\bmeta}) \xrightarrow{p}(\bpsi_0, \bar{\bmeta})$, where $ \bar\bmeta= (\bar \mu_D, \bar\mu_Y, \bar \pi) $ with either (i) $ \bar{\pi}= \pi_0 $ and $ \bar\mu_D =\mu_{D0}$; or (ii) $ \bar{\pi}= \pi_0$ and $ \bar\delta_Y/ \bar\delta_D=\beta_0({\x})$; or (iii)  $ \bar\mu_D= \mu_{D0},   \bar\mu_Y= \mu_{Y0}$, where $\bar\delta_C= \bar\mu_C(1,1,\x) -  \bar\mu_C(1,0,\x)-  \bar\mu_C(0,1,\x)+  \bar\mu_C(0,0,\x)$, $ C\in \{Y, D\} $.
	\\
	(b) For each $ \bpsi $ in an open subset of Euclidean space and each $ \bmeta $ in a metric space, let $ \varphi ( \bm o;\bpsi, \bmeta) $ be a measurable function such that the class of functions $ \{ \varphi ( \bm o;\bpsi, \bmeta) : \| \bpsi- \bpsi_0\|<\epsilon,  \| \mu_D- \bar{\mu}_D\|_2<\epsilon, \| \mu_Y- \bar{\mu}_Y\|_2<\epsilon,  \| \pi- \bar{\pi}\|_2<\epsilon \} $ is Donsker for some $ \epsilon>0 $, and such that $ E\|  \varphi ( \bmO ;\bpsi, \bmeta) -  \varphi ( \bmO;\bpsi_0, \bar{\bmeta})  \|^2\rightarrow 0 $ as $ (\bpsi, \bmeta)\rightarrow (\bpsi_0,  \bar{\bmeta}) $.  The maps $ \bpsi \mapsto E \{ \varphi ( \bmO ;\bpsi, \bmeta)\}$ are differentiable at $ \bpsi_0 $, uniformly in $ \bmeta $ in a neighborhood of $ \bar{\bmeta} $ with nonsingular derivative matrices $ M_{\bpsi_0, \bmeta} \rightarrow M_{\bpsi_0, \bar{\bmeta}} $. 
\end{assumption}
Assumption \ref{assump: semi}(a) describes the multiple robustness of our estimator. Assumption  \ref{assump: semi}(b) is  standard for $ Z $-estimators \citep[Chapter 5.4]{vanderVaart:2000book}.

\begin{theorem} \label{theo: 2}
	Under Assumptions \ref{assum: 1}-\ref{assump: semi},  $\hat \bpsi $ is consistent with rate of convergence 
	\begin{align}
		&	\|\hat{\bpsi}- \bpsi_0 \| =\nonumber\\
		&	O_p\left(n^{-1/2}	+ \| \hat{\pi}- \pi_0\|_2 \big(\| \hat{\mu}_Y- \mu_{Y0}\|_2 + \| \hat{\mu}_D- \mu_{D0}\|_2 \big)
		+ \bigg\|  \frac{\hat{\delta}_Y}{ \hat{\delta}_D} -\beta_0(\X) \bigg\|_2 \| \hat{\mu}_D - \mu_{D0}\|_2\right). \nonumber
	\end{align} 
	Suppose further that 
	\[
	\| \hat{\pi}- \pi_0\|_2 \big(\| \hat{\mu}_Y- \mu_{Y0}\|_2 + \| \hat{\mu}_D- \mu_{D0}\|_2 \big)
	+ \bigg\|  \frac{\hat{\delta}_Y}{ \hat{\delta}_D} -\beta_0(\X) \bigg\|_2 \| \hat{\mu}_D - \mu_{D0}\|_2= o_p(n^{-1/2}),
	\]
	then $ \hat{\bpsi} $ is asymptotically normal and semiparametric efficient, satisfying 
	\begin{align}
		\sqrt{n} ( \hat{\bpsi}- \bpsi_0) \xrightarrow{d} N\left(0, \ M_{\bpsi_0, \bmeta_0}^{-1} E\left\{  \varphi ( \bmO; \bpsi_0, \bmeta_0) \varphi ( \bmO; \bpsi_0, \bmeta_0)^T \right\} ( M_{\bpsi_0, \bmeta_0}^{-1} )^T\right). \label{eq: EIF}
	\end{align}
\end{theorem}
The first part of Theorem \ref{theo: 2} describes the convergence rate of $ \hat\bpsi $, which again indicates the multiple robustness of our estimator. That is, $\hat \bpsi $  is consistent provided that (i) either one of $ \hat{\pi} $ or $ (\hat{\mu}_Y, \hat{\mu}_D) $ is consistent, and (ii) either one of $ \hat{\delta}_Y/ \hat{\delta}_D $ or $\hat \mu_D $ is consistent. The multiple robustness property is important in practice, because nuisance parameters such as $ \pi(t, z, {\x} ) $ and $ \mu_D({\x}) $ may be easier to estimate than the outcome model $ \mu_Y({\x}) $. When all the nuisance parameters are consistently estimated, we can still benefit from using the semiparametric methods, in that even the nuisance parameters are estimated at slower rates, $ \hat{\bpsi} $ can still have fast convergence rate. For example, if all the nuisance parameters are estimated at $ n^{-1/4} $ rates, then $ \hat{\bpsi} $ can still achieve fast $ n^{-1/2} $ rate. The second part of Theorem \ref{theo: 2} says that if the nuisance parameters are consistently estimated with fast rates, for example, if they are estimated using parametric methods,  then their variance contributions  are negligible, and $ \hat{\bpsi} $  achieves the semiparametric efficiency bound.  

When (\ref{eq: EIF}) holds, a plug-in variance estimator for $ \sqrt{n} \hat{\bpsi} $ can be easily constructed as 
\begin{align}
	\hat{M}^{-1} \left[\frac{1}{n} \sum_{i=1}^{n}\varphi ( \bmO_i; \hat\bpsi, \hat\bmeta)  \varphi ( \bmO_i; \hat\bpsi, \hat\bmeta)^T\right] (\hat{M}^{-1})^T, \qquad  \hat{M}=\frac{1}{n}\sum_{i=1}^{n}\frac{ \partial \varphi (\bmO_i; \bpsi, \hat{\bmeta} )}{ \partial \bpsi}\bigg|_{ \bpsi= \hat\bpsi}, \nonumber
\end{align}
based on which we can perform hypothesis testing and construct confidence intervals. Even if (\ref{eq: EIF}) does not hold, when Assumption \ref{assump: semi} is true and  parametric methods are used to estimate all the nuisance parameters, inference using the bootstrap would still be valid, for example, even when $ \mu_Y(\x) $ is misspecified; see Section \ref{sec: simu} for empirical results. Furthermore, if one is worried about possible serial correlation among multiple measurements for an individual, then one can use the block bootstrap, which preserves the correlation by randomly sampling each individual together with all her measurements \citep{Shao:1995bootstrap}.

\section{Two-Sample Instrumented DID}
\label{sec: two-sample}
In some applications, it is hard to collect the exposure and outcome variables for the same individual, especially when the outcome is defined to reflect a delayed treatment effect. For instance, in the smoking and lung cancer example  in Section \ref{sec: intro}, the outcome of interest is lung cancer mortality after 35 years and it is infeasible to follow the same individuals for 35 years. Motivated from \cite{angrist1992effect, Angrist1995}'s influential two-sample standard IV analysis, we extend the instrumented DID to a two-sample design.

Suppose there are $ n_a $ i.i.d. realizations of $( T_{a}, Z_{a}, D_{a}, Y_{a} )$ from one sample, and $ n_b $ i.i.d.  realizations of  $( T_{b}, Z_{b}, D_{b}, Y_{b}  )$ from another sample. These two samples are independent of each other and we never observe $ D_{a} $ and $ Y_b $. We write the observed data as $( T_{ai}, Z_{ai}, Y_{ai},  i=1,\dots, n_a)$ and $( T_{bi}, Z_{bi}, D_{bi},  i=1,\dots, n_b )$, which are respectively referred to as the outcome dataset and the exposure dataset.  Let $ \delta_{Ya}, \hat\delta_{Ya},  \delta_{Db},\hat\delta_{Db}, \hat\mu_{Ya}(t, z),\hat\mu_{Db}(t,z)  $ be as defined in (\ref{eq: iv for trend, Wald}) and (\ref{eq: Wald estimator}) but evaluated correspondingly using the outcome dataset and exposure dataset. 	Suppose that Assumptions \ref{assum: 1}-\ref{assump: iv for trend} hold for the data generating processes in both datasets, and  $ E(Y_{a}|T_{a}, Z_{a}) =E(Y_{b}|T_{b}, Z_{b}) $,  $E(D_{a}|T_{a}, Z_{a}) =E(D_{b}|T_{b}, Z_{b}) $, then the average treatment effect  is identified by $ \beta_0= \delta_{Ya}/\delta_{Db}.  $
Analogously, the two-sample instrumented DID Wald estimator is obtained as $ 	\hat{\beta}_{\rm TS} = \hat\delta_{Ya}/\hat\delta_{Db}.  $  In Theorem S2 of the supplementary materials, we  establish the consistency and asymptotic normality of $  	\hat{\beta}_{\rm TS}  $ and provide a consistent variance estimator. Both 
$ \hat\beta_{\rm TS} $ and its variance estimator  can be conveniently calculated based on solely summary statistics  $\hat\mu_{Ya}(t, z)$ and $\hat\mu_{Db}(t, z)$ and their standard errors.

\section{Measure of Weak Identification}
\label{sec: weak}
Even when all the identification assumptions hold, estimation and  inference using the instrumented DID may still be unreliable under weak identification; see  \cite{Stock:2002aa} for a survey of weak identification in the standard IV setting. Different from the standard IV, weak identification for the instrumented DID   
arises when  trends in exposure for $ Z=0 $ and $ Z=1 $ are near-parallel. In this section, we develop a measure of weak identification tailored for the instrumented DID to serve as useful diagnostic checks.

%It can be  that $ \beta $ in \eqref{eq: iv for trend, Wald} is the same as the $\beta$ from a conventional two-stage least squares regression: $Y= \gamma_0+\beta D+\gamma_Z Z+\gamma_T T+ \epsilon_Y, D= \lambda_0+ \lambda_{TZ} ZT+\lambda_Z Z+\lambda_T T +\epsilon_D $, where $Z$ and $T$ are the included exogenous variables, and the interaction $ZT$ is the IV. 
Consider first the case when there are no observed covariates.  We take the one-sample estimator $ \hat\beta $ as an example; the result for the two-sample estimator $ \hat{\beta}_{\rm TS} $ is similar. Notice that $ \hat \delta_Y $ and $ \hat  \delta_D $ can be respectively obtained from fitting a saturated model of $ Y $ or $ D $ on $1,  ZT, Z$ and $ T $, where $ ZT  $ is the interaction term. Let $ \bm R $ be the $ n $-dimensional vector of residuals from regressing $ZT$ on $ 1, Z $ and $ T $. By using the Frisch-Waugh-Lovell theorem \citep{Davidson:1993book, Wang:1998aa},   $ \hat\beta $ in (\ref{eq: Wald estimator}) can be equivalently formulated as
\begin{align}
	\hat{\beta}=\frac{\hat \delta_Y}{\hat \delta_D} =\frac{( \bm{R}^T \bm R)^{-1}\bm{R}^T \bmY}{( \bm{R}^T \bm R)^{-1}\bm{R}^T \bmD}= \frac{\D^T H_{R} \Y }{\D^T H_R \D}, \nonumber
\end{align}
where $\bmD^T= (D_1 ,\dots, D_n), \bmY^T= (Y_1 ,\dots, Y_n)$, $ H_R = \bm{R}(\bm{R}^T\bm{R})^{-1} \bm{R}^T$ is the hat matrix. Interestingly, the above formula indicates that $ \hat{\beta} $ can be alternatively obtained from a conventional two-stage least squares: the exposure $ \bmD $ is first regressed on $ \bm R $ (first-stage regression) and the outcome $ \bm Y $ is then regressed on the predicted values from the first-stage regression. This provides a perception that $ Z $ as an IV for DID is equivalent to using $ ZT  $ as the standard IV while further controlling for 1, $ Z $  and $T $.  Hence, the concentration parameter of $ ZT $ as the standard IV (controlling for 1, $ Z $  and $T $) serves here as a measure of weak identification using $ Z $ as the IV for DID. Specifically, this measure is defined as $ 	{\kappa}^2=\delta_D^2 \bm{R}^T\bm{R}/ \sigma_\epsilon^2, $
where $ \delta_D$ is defined in Proposition \ref{prop: iv for trend, Wald}, $ \sigma_{\epsilon}^2 $ is the population residual variance from the first-stage regression. Heuristically, $ \kappa^2 $ increases if we have a larger sample size $ n $, larger $ \delta_D^2$, or a larger limit of $  \bm{R}^T\bm{R}/n $. For the usual inference based on normal approximation to be accurate, $\kappa^2 $ must be large. 

A commonly used estimate of  $\kappa^2 $ is the F  statistic from the first-stage regression. When only summary-data are available, i.e., only $ \hat\delta_D $ and its standard error are available, one can also use the squared z-score as an estimate of $ \kappa^2 $, where the z-score is the ratio of $ \hat\delta_D $ to its standard error. When there are observed covariates, a measure  of weak identification can also be easily calculated  by defining $ \bm R $  as the vector of residuals from regressing $ ZT$ on $ 1, Z, T, \X$. 
We follow  \cite{Stock:2005aa} and recommend checking to make sure that an estimated $\kappa^2 $ is larger than 10 before applying the inference methods in Sections \ref{sec: semi} and \ref{sec: two-sample}. 

%Otherwise, the usual inference methods may suffer from bias due to weak identification in the sense that the usual nominal 5\% Wald test of the hypothesis $ H_0: \beta=\beta_0 $ may have actual size exceeding 15\%. In this case, inference methods that are robust to weak identification should be applied, see \cite{Stock:2002aa} for a survey.

%On the other hand, weak identification also makes the IV for trend method more susceptible to bias arising from possible violations of the other assumptions.  For example, as derived in Section 1 of the supplementary materials, if the treatment effect changes over time so that Assumption \ref{assump: iv for trend}(c) does not hold, then there may be a bias 
%\[
%\frac{\delta_Y}{\delta_D}-  E(Y_1^{(1z)}- Y_1^{(0z)})= \frac{E( D_0^{(1)}- D_0^{(0)})}{\delta_D} \big\{ E(Y_1^{(1z)}- Y_1^{(0z)})- E(Y_0^{(1z)}- Y_0^{(0z)})\big\}. 
%\]
%Hence, any non-zero value in the numerator due to violations of Assumption \ref{assump: iv for trend}(c) will be amplified by a small denominator $ \delta_D $, resulting in a possibly large bias. Therefore, another measure of weak identification developed from a sensitivity analysis perspective is simply $| \hat{\delta}_D |$. If  $ |\hat\delta_D| $ is large, the IV for trend method is less sensitive to possible violations of the other assumptions. See  \cite{Wang:2018ab} for discussion under the standard IV setting.

\section{Simulations}
\label{sec: simu}

In this section, we conduct simulation studies to evaluate the finite sample performance of the proposed instrumented DID (iDID) method using two cases. For case 1, $ P(Z=1|X, U_0, U_1)=0.5 $; for case 2, $ P(Z=1|X, U_0, U_1)= \exp(0.5X)/(1+\exp(0.5X))$. The other variables are  from the same data generating process for the two cases, specifically, $ T\sim \text{Binom}(0.5) $, $ X\sim N(0,1) $, $ U_t\sim t+TN(0,1, (-1, 1)) $,  $ \epsilon_t\sim N(0,1) $,  $ P(D_t=1| U_0, U_1, X, Z) = (Z+1)U_t/8+0.5$, $ Y_t= (1+X)D_t+2+2U_t+Z+X+\epsilon_t $, for $ t=0,1 $, where $ TN(0,1, (-1,1)) $ denotes a truncated normal distribution with mean 0, variance 1, and support $ (-1,1) $. We simulate $ n=10^5 $ random samples from $ (T, Z, X,  D_0, D_1, Y_0, Y_1) $ and let $ D=TD_1+(1-T)D_0, Y=TY_1+(1-T)Y_0 $. The observed data is $ (Z_i, X_i, T_i, D_i, Y_i, i=1,\dots, n )$.

Under case 1, Assumptions \ref{assum: 1}-\ref{assump: iv for trend} hold with or without $ X $,  and thus  both the Wald estimator $ \hat{\beta} $ in (\ref{eq: Wald estimator}) and the semiparametric estimator $ \hat\bpsi $ in (\ref{eq: semi estimator}) using $ Z $ as the IV for DID are valid. Under case 2, Assumptions  \ref{assum: 1}-\ref{assump: iv for trend} hold only when conditioning on  $ X $, and thus the semiparametric estimator $ \hat\bpsi $  is valid, while the Wald estimator $ \hat{\beta} $ is not valid. We consider two working models for the semiparametric iDID method, a constant treatment effect working model (i.e., $ \beta(\bmv; \bpsi) = \psi$) and a linear treatment effect working model  (i.e., $ \beta(\bmv; \bpsi) = \psi_1+\psi_2 x $, with $ \bmV=\X $). The true values of $ \beta, \psi, \psi_1, \psi_2 $ are all equal to 1 because $ E(Y_t^{(1)}- Y_t^{(0)})=1$ and $ E(Y_t^{(1)}- Y_t^{(0)}|X)=1+X$. The weight function $ w(\bmv) $ in (\ref{eqn: obj}) is set to be 1.

We  also examine the effect of model misspecification for the semiparametric iDID estimators. Note that in cases 1-2, the functional forms of the nuisance parameters are
\begin{align}
	\pi (t, z, x) = 1/4 ~ \text{(for case 1)}, \quad& \pi (t, z, x) =  \frac{\{\exp(x)\}^z}{2\{1+\exp(x)\}} ~ \text{(for case 2)},\nonumber\\
	\mu_D(t, z, x)= (z+1)t/8+0.5, \quad & \mu_Y(t, z, x)= (1+x) \{  (z+1)t/8+0.5 \}+ 2+ t+z+x.  \nonumber
\end{align} 
Therefore, the correct model we fit for $ \pi(t, z, x) $ is the product of two logistic models, one for $ P(Z=z|X=x, T=t) $ and one for $ P(T=t|X=x) $; the correct models we fit for $ \mu_D(t, z, x), \mu_Y(t, z, x) $ are linear models with all the main effects and interactions among $ t, z, x $. The misspecified model we fit for $ \mu_D (t, z, x)$ is a logistic model; the misspecified models we fit for  $ \mu_Y(t,z,x), \pi(t,z,x) $ are respectively replacing $ x $ in the correct models with $ \exp(x/2) $, which is similar to the covariate transformation in \cite{Kang:2007aa}. 

We compare with  two other methods,  direct treated-vs.-control outcome comparison using ordinary least squares (OLS) and the standard IV method using $ Z $ as the IV. Direct outcome comparison is invalid because of the unmeasured confounder $ U_t $; the standard IV method is also invalid due to the direct effect of $ Z $ on the outcome. The standard IV method is implemented using the R package \textsf{ivpack} \citep{ivpack}. Tables \ref{tb: table1}-\ref{tb: table2} show the simulation results based on 1000 repetitions. Specifically, Tables \ref{tb: table1}-\ref{tb: table2}  include (i) the simulation average bias and standard deviation (SD) of each estimator; (ii) the median of standard errors (SEs), which are calculated according to Equation (S3) in the supplementary materials for the Wald estimator, using the percentile bootstrap with 200 bootstrap iterations for the semiparametric estimators; (iii) simulation coverage probability (CP) of 95\% confidence intervals. For case 1, because $ \pi $ is always correctly specified, we only examine the effects of misspecifying $ \mu_D $ and $ \mu_Y $.

\begin{table}
	\caption{ Simulation results for case 1 based on 1,000 repetitions. In the third column, $ \hat{\beta} $ denotes the Wald estimator (\ref{eq: Wald estimator}), $ \hat{\psi} $ denotes the semiparametric estimator  (\ref{eq: semi estimator}) with a constant working model $ \beta(x; \bpsi)=\psi $, and $ \hat{\psi}_1, \hat{\psi}_2 $ denote that  with a linear working model $ \beta(x; \bpsi)=\psi_1+\psi_2x $.	The underlined scenarios are the ones that Theorem \ref{theo: 2} predicts the semiparametric estimators to be consistent. ($n=10^5$, the true values of  $\beta, \psi, \psi_{1}, \psi_2$ are all equal to 1). \label{tb: table1}}
	\centering
	\begin{tabular}{llcccccc} \\ [-2ex] \hline
		Correct Model                       & Method      & Estimator && Bias   & SD    & SE    & CP    \\ [0.5ex] \cline{1-3} \cline{5-8} \\ [-2ex]
		& OLS          &           && 0.906  & 0.015 & 0.016 & 0 \\
		& Standard IV           &           && 16.049 & 0.801 & 0.790 & 0 \\
		%	& calendar IV  &           && 10.667 & 0.176 & 0.182 & 0 \\
		&              &           &        &       &       &       &            \\[-1.5ex]
		& iDID & $ \hat\beta $      && -0.002  & 0.226 & 0.226 & 0.956 \\
		&              &           &        &       &       &       \\[-1.5ex]
		\multirow{3}{*}{(\underline{$\pi, \mu_D, \mu_Y$)}} &iDID & $\hat{\psi}$      && -0.010  & 0.150 & 0.150 & 0.952 \\
		& iDID  & $\hat{\psi}_1$       && -0.010 & 0.150 & 0.149 & 0.945 \\
		& iDID & $\hat{\psi}_2$       && -0.010  & 0.150 & 0.156 & 0.962 \\
		&              &           &        &       &&       &       \\[-1.5ex]
		\multirow{3}{*}{($\pi, \mu_Y)$}        & iDID & $\hat{\psi}$       && -0.790  & 0.032 & 0.032 & 0 \\
		& iDID  & $\hat{\psi}_1$       && -0.790  & 0.032 & 0.032 & 0 \\
		& iDID  & $\hat{\psi}_2$       && -0.789 & 0.034 & 0.034 & 0 \\
		&              &           &        &       &      & &       \\[-1.5ex]
		\multirow{3}{*}{\underline{($\pi, \mu_D$)}}        & iDID  & $\hat{\psi}$       && -0.009  & 0.160 & 0.160 & 0.953 \\
		&iDID  & $\hat{\psi}_1$       && -0.009  & 0.160 & 0.161 & 0.952 \\
		&iDID & $\hat{\psi}_2$       && -0.010  & 0.201 & 0.209 & 0.962 \\
		&              &           &        &       &       &&       \\[-1.5ex]
		\multirow{3}{*}{($\pi$)}               &iDID  & $\hat{\psi}$       && -0.789  & 0.034 & 0.034 & 0 \\
		& iDID & $\hat{\psi}_1$       && -0.789 & 0.034 & 0.034 & 0 \\
		&iDID  & $\hat{\psi}_2$       && -0.789 & 0.043 & 0.044 & 0.001 \\[0.5ex] \hline
	\end{tabular}
\end{table}

\begin{table}
	\caption{Simulation results for case 2 based on 1,000 repetitions. In the third column, $ \hat{\beta} $ denotes the Wald estimator (\ref{eq: Wald estimator}), $ \hat{\psi} $ denotes the semiparametric estimator  (\ref{eq: semi estimator}) with a constant working model $ \beta(x; \bpsi)=\psi $, and $ \hat{\psi}_1, \hat{\psi}_2 $ denote that  with a  linear working model $ \beta(x; \bpsi)=\psi_1+\psi_2x $.	The underlined scenarios are the ones that Theorem \ref{theo: 2} predicts the semiparametric estimators to be consistent. ($n=10^5$, the true values of  $\beta, \psi, \psi_{1}, \psi_2$ are all equal to 1). \label{tb: table2}}
	\centering
	\begin{tabular}{llcccccc} \\ [-2ex] \hline
		Correct Model                       & Method      & Estimator && Bias   & SD    & SE    & CP    \\ [0.5ex] \cline{1-3} \cline{5-8} \\ [-2ex]
		& OLS          &  &  & 1.658   & 0.018 & 0.018 & 0     \\
		& Standard IV           &  &  & -17.122 & 0.639 & 0.622 & 0     \\
		%	& calendar IV  &  &  & 10.604  & 0.179 & 0.180 & 0     \\
		&              &  &  &         &       &       &      \\[-1.5ex]
		&iDID & $ \hat\beta $  &  & -0.630  & 0.246 & 0.249 & 0.272 \\
		&              &  &  &         &       &       &    \\[-1.5ex]
		\multirow{3}{*}{\underline{($\pi, \mu_D, \mu_Y$)}} 
		& iDID  & $\hat{\psi}$ &  & -0.018  & 0.205 & 0.204 & 0.960 \\
		& iDID  & $\hat{\psi}_1$ &  & -0.018  & 0.205 & 0.204 & 0.952 \\
		& iDID  & $\hat{\psi}_2$ &  & -0.003  & 0.228 & 0.224 & 0.955 \\
		&              &  &  &         &       &       &    \\[-1.5ex]
		\multirow{3}{*}{(\underline{$\mu_D, \mu_Y$})} 
		&iDID  & $\hat{\psi}$ &  & -0.019  & 0.268 & 0.219 & 0.952 \\
		& iDID  & $\hat{\psi}_1$ &  & -0.019  & 0.268 & 0.218 & 0.959 \\
		&iDID  & $\hat{\psi}_2$ &  & -0.001  & 0.854 & 0.318 & 0.967 \\
		&              &  &  &         &       &       &     \\[-1.5ex]
		\multirow{3}{*}{($\pi, \mu_Y$)} 
		& iDID & $\hat{\psi}$ &  & -0.764  & 0.049 & 0.049 & 0 \\
		& iDID  & $\hat{\psi}_1$ &  & -0.764  & 0.049 & 0.049 & 0 \\
		&iDID & $\hat{\psi}_2$ &  & -0.760  & 0.057 & 0.056 & 0.001 \\
		&              &  &  &         &       &       &      \\[-1.5ex]
		\multirow{3}{*}{\underline{($\pi, \mu_D$)}} 
		&iDID & $\hat{\psi}$ &  & -0.022  & 0.212 & 0.215 & 0.960 \\
		&iDID& $\hat{\psi}_1$ &  & -0.022  & 0.212 & 0.214 & 0.957 \\
		&iDID & $\hat{\psi}_2$ &  & -0.018  & 0.277 & 0.282 & 0.956 \\
		&              &  &  &         &       &       &       \\[-1.5ex]
		\multirow{3}{*}{($\mu_Y$)} 
		& iDID & $\hat{\psi}$ &  & -0.763  & 0.072 & 0.055 & 0.005 \\
		& iDID & $\hat{\psi}_1$ &  & -0.763  & 0.072 & 0.055 & 0.006 \\
		& iDID& $\hat{\psi}_2$ &  & -0.752  & 0.243 & 0.096 & 0.048 \\
		&              &  &  &         &       &       &       \\[-1.5ex]
		\multirow{3}{*}{($\mu_D$)} 
		& iDID & $\hat{\psi}$ &  & -0.152  & 0.959 & 0.323 & 0.949 \\
		&iDID & $\hat{\psi}_1$ &  & -0.151  & 0.958 & 0.324 & 0.952 \\
		& iDID & $\hat{\psi}_2$ &  & -0.202  & 4.305 & 0.901 & 0.976 \\
		&              &  &  &         &       &       &      \\[-1.5ex]
		\multirow{3}{*}{($\pi$)} 
		& iDID & $\hat{\psi}$ &  & -0.764  & 0.051 & 0.051 & 0.001 \\
		&iDID & $\hat{\psi}_1$ &  & -0.764  & 0.051 & 0.051 & 0 \\
		& iDID & $\hat{\psi}_2$ &  & -0.762  & 0.068 & 0.067 & 0.001 \\
		&              &  &  &         &       &       &     \\[-1.5ex]
		\multirow{3}{*}{(none)} 
		&iDID& $\hat{\psi}$ &  & -0.790  & 0.275 & 0.075 & 0.040 \\
		& iDID & $\hat{\psi}_1$ &  & -0.790  & 0.275 & 0.076 & 0.041 \\
		& iDID & $\hat{\psi}_2$ &  & -0.779  & 1.262 & 0.212 & 0.230\\[0.5ex] \hline
	\end{tabular}
\end{table}

The following is a summary based on the results in Tables \ref{tb: table1}-\ref{tb: table2}. First, OLS and standard IV have large bias due to violations of their assumptions. The iDID Wald estimator $ \hat{\beta} $ shows negligible bias and adequate coverage probability in case 1, but is biased  in case 2, which is anticipated and is due to the correlation between $ Z $ and $ X $. In both cases, the semiparametric iDID estimators exhibit negligible bias and adequate coverage probabilities when $ (\pi, \mu_D, \mu_Y), (\pi, \mu_D), (\mu_D, \mu_Y) $ are correctly specified, which supports the multiple robustness property. Notice that in the considered simulation setups, even when all the nuisance parameters are misspecified or with Assumption \ref{assump: iv for trend} being violated, the iDID semiparametric and  Wald  estimators still have smaller bias compared with the other methods. Second, when $ \pi $ is misspecified, the semiparametric estimators may be unstable because $ \pi $ appears in the denominator and thus the SD can be inflated if some $ \hat\pi $ are close to zero. Nonetheless, the average bias is still small and coverage probability is adequate (larger than 0.95), which agrees with our theory. In the other underlined scenarios that our theory predicts the semiparametric estimators to be consistent, all SEs are close to the simulation SDs, even when part of the nuisance parameters is misspecified. Lastly, compared within the semiparametric iDID estimators in the underlined scenarios, the set of estimators with all the nuisance parameters correctly specified have the smallest simulation SDs, which  agrees with our efficiency results in Theorem \ref{theo: 2}.

\section{Application}
\label{sec: data}

We apply the proposed method to  analyze the effect of cigarette smoking on lung cancer mortality.  Given the lag between smoking exposure and lung cancer mortality, we adopt the two-sample instrumented DID design. Our analysis is based upon two datasets arranged by 10-year birth cohort: the 1970 National Health Interview Survey (NHIS) for nationally representative estimates of smoking prevalence \citep{nhis-1970}, and the US Centers for Disease Control and Prevention's (CDC) Wide-ranging ONline Data for Epidemiologic Research (WONDER) system for estimates of national lung cancer (ICD-8/9: 162; ICD-10: C33-C34) mortality rates \citep{cdc1968-1978,cdc1979-1998,cdc1999-2016}. 
Only the 1970 NHIS is used because it is the first NHIS that  records the initiation and cessation time of smoking such that a longitudinal structure is available. We closely follow the approach taken by   \citet[Chapter 3]{Tolley:1991} to calculate the smoking prevalence rates. %based on responses to four questions: ``Have you smoked at least 100 cigarettes during your entire life?'', ``Do you smoke cigarettes now?'', ``how long has it been since you smoked cigarettes fairly regularly?'',  and ``How old were you when you first started smoking cigarettes fairly regularly?''.

\begin{figure}[t]
	\centering
	\includegraphics[scale=0.7]{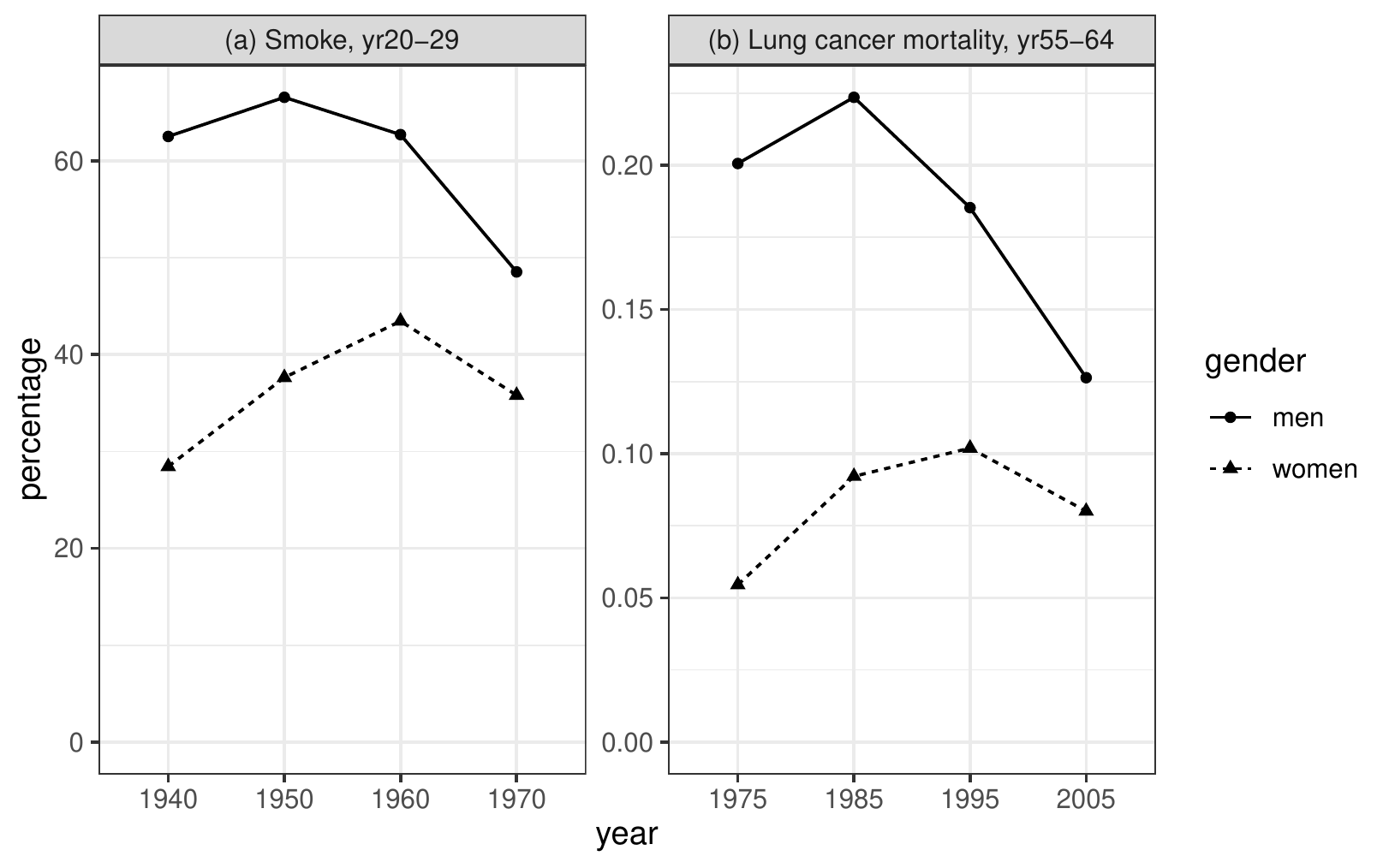}
	\caption{\label{fig: data} Changes in prevalence of cigarette smoking for men and women aged 20-29, lung cancer mortality rates for men and women aged 55-64 among four successive 10-year birth cohorts: 1911-1920, 1921-1930, 1931-1940, 1941-1950.}
\end{figure}

Based on the data availability, we focus on four successive 10-year birth cohorts: 1911-1920, 1921-1930, 1931-1940, 1941-1950, whose smoking prevalence is estimated respectively at year 1940, 1950, 1960, 1970 when they are at age 20-29, whose lung cancer mortality rates are estimated respectively at year 1975, 1985, 1995, 2005 when they are at age 55-64. Here, cohort of birth plays the role of time. Figure \ref{fig: data} shows the changes in prevalence of cigarette smoking among men and women aged 20-29, and the changes in lung cancer mortality rates 35 years later in the United States. From Figure \ref{fig: data}, we see that the trends in lung cancer mortality rates follow the trends in smoking prevalence, with a lag of 35 years, which provides evidence that smoking increases lung cancer mortality rate.

There have been many direct comparisons of the lung cancer mortality rates between smokers and non-smokers which have found higher rates among smokers \citep{iarc1986tobacco}. Additional studies that replicate direct comparisons of smokers and non-smokers may not add much evidence beyond the first comparison. It is argued in \cite{rosenbaum2010design} that ``in such a situation, it may be possible to find haphazard nudges that, at the margin, enable or discourage [the exposure]. ... These nudges may be biased in various ways, but there may be no reason for them to be consistently biased in the same direction, so similar estimates of effect from studies subject to different potential biases gradually reduce ambiguity about what part is effect and what part is bias.''  The instrumented DID is one such method that attempts to exploit the ``haphazard nudges'', i.e., the targeted tobacco advertising to women in the 1960s that led to a rapid increase in smoking among young women in a way that is presumably independent of other causes of lung cancer mortality. 

\begin{table}
	\caption{Two-sample iDID Wald estimates and their standard errors (in parentheses) using two successive birth cohorts (in \%). F-statistic is the squared z-score, $ \hat\beta_{\rm TS} $ defined in Section \ref{sec: two-sample} estimates the average treatment effect of smoking on lung cancer mortality. \label{tb: data} }
	\centering
	\begin{tabular}{lccc} \\ [-2ex] \hline
		\multirow{2}{*}{Birth Cohort} & 1911-1920 & 1921-1930 & 1931-1940 \\
		& 1921-1930 & 1931-1940 & 1941-1950\\ [0.5ex]
		\hline \\ [-1.5ex]
		F-statistic & 13.94 &47.28& 21.33\\
	%	$\hat \delta_D $& 5.1 & 9.7 & 6.5 \\ 
		$\hat \beta_{\rm TS} $ & 0.285 (0.089) & 0.497 (0.076) & 0.568 (0.127)    \\[0.5ex] \hline
	\end{tabular}
\end{table}

To quantitatively evaluate the effect of cigarette smoking on lung cancer mortality, we take gender -- a surrogate of whether each individual received encouragement (targeted tobacco advertising) or not -- as the IV for DID. Note that gender does not need to have a causal effect on smoking; as  proved in the supplementary materials,  it suffices that gender is correlated with smoking due to the encouragement from targeted tobacco advertising.  We consider two successive 10-year birth cohorts, setting the earlier birth cohort as $T=0$ and the later birth cohort as $T=1$. Gender is likely a valid IV for DID, as it clearly satisfies the trend relevance assumption, the lung cancer mortality rates for men and women would have evolved similarly had all subjects counterfactually not smoked, and there is no evident gender difference in the cancer-causing effects of cigarette smoking \citep{Patel:2004aa}. 

Table \ref{tb: data} summarizes (i) the F-statistic proposed in Section \ref{sec: weak} to measure weak identification; and (ii) the two-sample iDID Wald estimators  $ \hat\beta_{\rm TS} $ defined in Section \ref{sec: two-sample} and their standard errors defined in Equation (S5) of the supplementary materials.  More details on the application are also in the supplementary materials. 

From Table \ref{tb: data}, under the assumption that gender is a valid IV for DID and the treatment effect is stable over time, we found evidence that smoking leads to significantly higher  lung cancer mortality rates. Specifically, we find that smoking in one's 20s leads to an elevated annual lung cancer mortality rate at age 55-64, with the effect size ranging from 0.285\% to 0.568\%. This is of a similar magnitude as the findings in  \cite{Thun1982, Thun:2013aa}. Using different birth cohorts gives slightly different point estimates, but they are within two standard errors of each other.  Nonetheless, there is still concern about violating the stable treatment effect over time assumption (Assumption \ref{assump: iv for trend}(d)), possibly because the   cigarette design and composition have undergone changes that promote  deeper inhalation of smoke \citep{Thun:2013aa, Warren:2014aa}. In the supplementary materials (Section S3), we perform a sensitivity analysis and find that increasing risk of smoking over time does not explain away the observed treatment effect. 

%We generally recommend applying the IV for trend method when $|\hat \delta_D|$ is large, as larger $|\hat \delta_D|$ guarantees the estimation and inference being more reliable and usually leads to higher precision. 

%the increasing risk of smoking over time is also observed in other studies, and a plausible explanation is that

\section{Results and Discussion}
\label{sec: discussions}
In this paper, we have proposed a new method called instrumented DID that explicitly leverages exogenous randomness in  the exposure trends, and controls for unmeasured confounding in longitudinal or repeated cross sectional studies. The instrumented DID method evolves from two powerful natural experiment devices, the standard  IV and standard  DID, but is able to relax some of their most disputable assumptions.  Our motivation of assessing the causal effect by linking the change in outcome mean and the change in exposure rate is also related to the trend-in-trend design \citep{Ji2017} and etiologic mixed design \citep{lash2021modern}.

%which identifies the causal odds ratio under a structural logistic model. 

 %Reviews of existing methods for addressing unmeasured confounding in observational studies can be found in \cite{streeter2017adjusting} and \cite{Zhang:2018review}. 

In principle, any variable that satisfies Assumptions \ref{assump: iv for trend}(a)-(c) can be chosen as the IV for DID. Here, we list two common sources of the IV for DID: (i) administrative information, such as geographic region and insurance type; and  (ii) variables that are commonly used as standard IVs, such as physician preference, distance to care provider, and genetic variants -- see \cite{Baiocchi:2014aa} for more examples; as discussed in Section \ref{sec: iv for trend}, these variables are more likely to be an IV for DID compared  to being a standard IV, because  IVs for DID are allowed to have direct effects on the outcome.

%\section*{Acknowledgements}
%We gratefully acknowledge support from FWO-Vlaanderen Research Project
%``Sensitivity Analysis for Incomplete and Coarse Data'' and Belgian
%IUAP/PAI network ``Statistical Techniques and Modeling for Complex
%Substantive Questions with Complex Data''. We are very grateful to
%David Cox for sharing his insights with us and for very helpful
%comments on an earlier version of this manuscript.

%  Not included in the original version of this paper!

%  Here, we create the bibliographic entries manually, following the
%  journal style.  If you use this method or use natbib, PLEASE PAY
%  CAREFUL ATTENTION TO THE BIBLIOGRAPHIC STYLE IN A RECENT ISSUE OF
%  THE JOURNAL AND FOLLOW IT!  Failure to follow stylistic conventions
%  just lengthens the time spend copyediting your paper and hence its
%  position in the publication queue should it be accepted.

%  We greatly prefer that you incorporate the references for your
%  article into the body of the article as we have done here 
%  (you can use natbib or not as you choose) than use BiBTeX,
%  so that your article is self-contained in one file.
%  If you do use BiBTeX, please use the .bst file that comes with 
%  the distribution.

\vspace*{-8pt}

\bibliographystyle{apalike} 
\bibliography{reference}

\begin{thebibliography}{}

\bibitem[Abadie, 2003]{Abadie:2003aa}
Abadie, A. (2003).
\newblock Semiparametric instrumental variable estimation of treatment response
  models.
\newblock {\em Journal of Econometrics}, 113(2):231--263.

\bibitem[Abadie, 2005]{abadie2005semiparametric}
Abadie, A. (2005).
\newblock Semiparametric difference-in-differences estimators.
\newblock {\em The Review of Economic Studies}, 72(1):1--19.

\bibitem[Angrist et~al., 1996]{Angrist:1996aa}
Angrist, J.~D., Imbens, G.~W., and Rubin, D.~B. (1996).
\newblock Identification of causal effects using instrumental variables.
\newblock {\em Journal of the American Statistical Association},
  91(434):444--455.

\bibitem[Angrist and Krueger, 1992]{angrist1992effect}
Angrist, J.~D. and Krueger, A.~B. (1992).
\newblock The effect of age at school entry on educational attainment: an
  application of instrumental variables with moments from two samples.
\newblock {\em Journal of the American statistical Association},
  87(418):328--336.

\bibitem[Angrist and Krueger, 1995]{Angrist1995}
Angrist, J.~D. and Krueger, A.~B. (1995).
\newblock Split-sample instrumental variables estimates of the return to
  schooling.
\newblock {\em Journal of Business \& Economic Statistics}, 13(2):225--235.

\bibitem[Angrist and Pischke, 2008]{Angrist:2008aa}
Angrist, J.~D. and Pischke, J.-S. (2008).
\newblock {\em Mostly harmless econometrics: An empiricist's companion}.
\newblock Princeton University Press.

\bibitem[Bailar and Gornik, 1997]{Bailar:1997aa}
Bailar, J.~C. and Gornik, H.~L. (1997).
\newblock Cancer undefeated.
\newblock {\em New England Journal of Medicine}, 336(22):1569--1574.

\bibitem[Baiocchi et~al., 2014]{Baiocchi:2014aa}
Baiocchi, M., Cheng, J., and Small, D.~S. (2014).
\newblock Instrumental variable methods for causal inference.
\newblock {\em Statistics in Medicine}, 33(13):2297--2340.

\bibitem[Bickel et~al., 1993]{Bickel:1993book}
Bickel, P., Klaassen, C., Ritov, Y., and Wellner, J. (1993).
\newblock {\em Efficient and Adaptive Estimation for Semiparametric Models}.
\newblock Springer.

\bibitem[Burbank, 1972]{Burbank:1972aa}
Burbank, F. (1972).
\newblock {U.S.} lung cancer death rates begin to rise proportionately more
  rapidly for females than for males: A dose-response effect?
\newblock {\em Journal of Chronic Diseases}, 25(8):473--479.

\bibitem[Card and Krueger, 1994]{Card:1994aa}
Card, D. and Krueger, A.~B. (1994).
\newblock Minimum wages and employment: A case study of the fast food industry
  in new jersey and pennsylvania.
\newblock {\em American Economic Review}, 84:772--793.

\bibitem[CDC, 2000a]{cdc1968-1978}
CDC (2000a).
\newblock Centers for disease control and prevention, national center for
  health statistics. compressed mortality file 1968-1978. {CDC WONDER} online
  database, compiled from compressed mortality file {CMF} 1968-1988, series 20,
  no. 2{A}, 2000. accessed at http://wonder.cdc.gov/cmf-icd8.html on {A}ug 27,
  2020.

\bibitem[CDC, 2000b]{cdc1979-1998}
CDC (2000b).
\newblock Centers for disease control and prevention, national center for
  health statistics. compressed mortality file 1979-1998. {CDC WONDER} online
  database, compiled from compressed mortality file {CMF} 1979-1998, series 20,
  no. 2{A}, 2000 and {CMF} 1989-1998, series 20, no. 2{E}, 2003. accessed at
  http://wonder.cdc.gov/cmf-icd9.html on {A}ug 27, 2020.

\bibitem[CDC, 2016]{cdc1999-2016}
CDC (2016).
\newblock Centers for disease control and prevention, national center for
  health statistics. compressed mortality file 1999-2016 on cdc wonder online
  database, released june 2017. data are from the compressed mortality file
  1999-2016 series 20 no. 2{U}, 2016. accessed at
  http://wonder.cdc.gov/cmf-icd10.html on {A}ug 28, 2020.

\bibitem[Cui and Tchetgen~Tchetgen, 2021]{Cui:2020aa}
Cui, Y. and Tchetgen~Tchetgen, E. (2021).
\newblock A semiparametric instrumental variable approach to optimal treatment
  regimes under endogeneity.
\newblock {\em Journal of the American Statistical Association},
  116(533):162--173.

\bibitem[Davidson and MacKinnon, 1993]{Davidson:1993book}
Davidson, R. and MacKinnon, J.~G. (1993).
\newblock {\em Estimation and Inference in Econometrics}.
\newblock Oxford University Press.

\bibitem[de~Chaisemartin and D'Haultf{\OE}uille, 2017]{Chaisemartin:2017aa}
de~Chaisemartin, C. and D'Haultf{\OE}uille, X. (2017).
\newblock Fuzzy differences-in-differences.
\newblock {\em The Review of Economic Studies}, 85(2):999--1028.

\bibitem[Duflo, 2001]{duflo2001schooling}
Duflo, E. (2001).
\newblock Schooling and labor market consequences of school construction in
  indonesia: Evidence from an unusual policy experiment.
\newblock {\em American economic review}, 91(4):795--813.

\bibitem[Fogarty, 2020]{fogary2017avg}
Fogarty, C.~B. (2020).
\newblock Studentized sensitivity analysis for the sample average treatment
  effect in paired observational studies.
\newblock {\em Journal of the American Statistical Association},
  115(531):1518--1530.

\bibitem[Hern{\'a}n and Robins, 2006]{Hernan:2006aa}
Hern{\'a}n, M.~A. and Robins, J.~M. (2006).
\newblock Instruments for causal inference: An epidemiologist's dream?
\newblock {\em Epidemiology}, 17(4):360--372.

\bibitem[Hernan and Robins, 2020]{Hernan-Robins}
Hernan, M.~A. and Robins, J.~M. (2020).
\newblock {\em Causal Inference: What If}.
\newblock Boca Raton: Chapman \& Hall/CRC.

\bibitem[Imbens, 2003]{Imbens:2003}
Imbens, G.~W. (2003).
\newblock Sensitivity to exogeneity assumptions in program evaluation.
\newblock {\em The American Economic Review Papers and Proceedings},
  93(2):126--132.

\bibitem[{International Agency for Research on Cancer}, 1986]{iarc1986tobacco}
{International Agency for Research on Cancer} (1986).
\newblock {\em Tobacco smoking}, volume~38.
\newblock World Health Organization.

\bibitem[Ji et~al., 2017]{Ji2017}
Ji, X., Small, D.~S., Leonard, C.~E., and Hennessy, S. (2017).
\newblock The trend-in-trend research design for causal inference.
\newblock {\em Epidemiology}, 28(4):529--536.

\bibitem[Jiang and Small, 2014]{ivpack}
Jiang, Y. and Small, D.~S. (2014).
\newblock {\em ivpack: Instrumental Variable Estimation.}
\newblock R package version 1.2.

\bibitem[Kang and Schafer, 2007]{Kang:2007aa}
Kang, J. D.~Y. and Schafer, J.~L. (2007).
\newblock Demystifying double robustness: A comparison of alternative
  strategies for estimating a population mean from incomplete data.
\newblock {\em Statistical Science}, 22(4):523--539.

\bibitem[Kennedy et~al., 2019]{Kennedy:2019aa}
Kennedy, E.~H., Lorch, S., and Small, D.~S. (2019).
\newblock Robust causal inference with continuous instruments using the local
  instrumental variable curve.
\newblock {\em Journal of the Royal Statistical Society: Series B (Statistical
  Methodology)}, 81(1):121--143.

\bibitem[Lash et~al., 2021]{lash2021modern}
Lash, T.~L., VanderWeele, T.~J., Haneuse, S., and Rothman, K.~J. (2021).
\newblock {\em Modern epidemiology}, volume~4.
\newblock Wolters Kluwer Health.

\bibitem[Lawlor et~al., 2004]{Lawlor:2004aa}
Lawlor, D.~A., Davey~Smith, G., Kundu, D., Bruckdorfer, K.~R., and Ebrahim, S.
  (2004).
\newblock Those confounded vitamins: what can we learn from the differences
  between observational versus randomised trial evidence?
\newblock {\em Lancet}, 363(9422):1724--1727.

\bibitem[Meigs, 1977]{Meigs:1977aa}
Meigs, J.~W. (1977).
\newblock Epidemic lung cancer in women.
\newblock {\em JAMA}, 238(10):1055--1055.

\bibitem[{National Health Interview Survey}, 1970]{nhis-1970}
{National Health Interview Survey} (1970).
\newblock Accessed at
  ftp://ftp.cdc.gov/pub/health\_statistics/nchs/datasets/nhis/1970 on {A}ug 31,
  2020.

\bibitem[Neyman, 1923]{Neyman:1923a}
Neyman, J. (1923).
\newblock On the application of probability theory to agricultural experiments.
  essay on principles. section 9.
\newblock {\em Statistical Science}, 5(4):465--472. Trans. Dorota M. Dabrowska
  and Terence P. Speed (1990).

\bibitem[Ogburn et~al., 2015]{Ogburn:2015aa}
Ogburn, E.~L., Rotnitzky, A., and Robins, J.~M. (2015).
\newblock Doubly robust estimation of the local average treatment effect curve.
\newblock {\em Journal of the Royal Statistical Society: Series B (Statistical
  Methodology)}, 77(2):373--396.

\bibitem[Patel et~al., 2004]{Patel:2004aa}
Patel, J.~D., Bach, P.~B., and Kris, M.~G. (2004).
\newblock Lung cancer in us women: A contemporary epidemic.
\newblock {\em JAMA}, 291(14):1763--1768.

\bibitem[Pierce and Gilpin, 1995]{pierce1995historical}
Pierce, J.~P. and Gilpin, E.~A. (1995).
\newblock A historical analysis of tobacco marketing and the uptake of smoking
  by youth in the united states: 1890--1977.
\newblock {\em Health Psychology}, 14(6):500.

\bibitem[Rosenbaum, 1987]{Rosenbaum:1987aa}
Rosenbaum, P.~R. (1987).
\newblock The role of a second control group in an observational study.
\newblock {\em Statist. Sci.}, 2(3):292--306.

\bibitem[Rosenbaum, 2010]{rosenbaum2010design}
Rosenbaum, P.~R. (2010).
\newblock {\em Design of observational studies}.
\newblock Springer.

\bibitem[Rubin, 1974]{Rubin:1974}
Rubin, D.~B. (1974).
\newblock Estimating causal effects of treatments in randomized and
  nonrandomized studies.
\newblock {\em Journal of Educational Psychology}, 6(5):688--701.

\bibitem[Rubin, 1978]{Rubin:1978aa}
Rubin, D.~B. (1978).
\newblock Bayesian inference for causal effects: The role of randomization.
\newblock {\em Annals of Statistics}, 6(1):34--58.

\bibitem[Rubin, 1990]{Rubin:1990aa}
Rubin, D.~B. (1990).
\newblock Comment: Neyman (1923) and causal inference in experiments and
  observational studies.
\newblock {\em Statistical Science}, 5(4):472--480.

\bibitem[Rutter, 2007]{Rutter:2007aa}
Rutter, M. (2007).
\newblock Identifying the environmental causes of disease: How should we decide
  what to believe and when to take action?
\newblock Report Synopsis. Academy of Medical Sciences.

\bibitem[Shao and Tu, 2012]{Shao:1995bootstrap}
Shao, J. and Tu, D. (2012).
\newblock {\em The Jackknife and Bootstrap}.
\newblock Springer.

\bibitem[Shi et~al., 2020]{Shi:2020aa}
Shi, X., Miao, W., Nelson, J.~C., and Tchetgen~Tchetgen, E.~J. (2020).
\newblock Multiply robust causal inference with double-negative control
  adjustment for categorical unmeasured confounding.
\newblock {\em Journal of the Royal Statistical Society: Series B (Statistical
  Methodology)}, 82(2):521--540.

\bibitem[Small, 2007]{small2007sensitivity}
Small, D.~S. (2007).
\newblock Sensitivity analysis for instrumental variables regression with
  overidentifying restrictions.
\newblock {\em Journal of the American Statistical Association},
  102(479):1049--1058.

\bibitem[Stock and Yogo, 2005]{Stock:2005aa}
Stock, J. and Yogo, M. (2005).
\newblock Testing for weak instruments in linear {IV} regression.
\newblock {\em Andrews DWK Identification and Inference for Econometric Models.
  New York: Cambridge University Press}, pages 80--108.

\bibitem[Stock et~al., 2002]{Stock:2002aa}
Stock, J.~H., Wright, J.~H., and Yogo, M. (2002).
\newblock A survey of weak instruments and weak identification in generalized
  method of moments.
\newblock {\em Journal of Business \& Economic Statistics}, 20(4):518--529.

\bibitem[Tan, 2006]{tan2006regression}
Tan, Z. (2006).
\newblock Regression and weighting methods for causal inference using
  instrumental variables.
\newblock {\em Journal of the American Statistical Association},
  101(476):1607--1618.

\bibitem[Thun et~al., 1982]{Thun1982}
Thun, J.~M., Day-Lally, C., Myers, G.~D., Calle, E.~E., Flanders, W.~D., Zhu,
  B.-P., and et~al. (1982).
\newblock Trends in tobacco smoking and mortality from cigarette use in cancer
  prevention studies {I}(1959-1965) and {II}(1982-1988).
\newblock Changes in cigarette-related disease risks and their implication for
  prevention and control: smoking and tobacco control monograph 8.

\bibitem[Thun et~al., 2013]{Thun:2013aa}
Thun, M.~J., Carter, B.~D., Feskanich, D., Freedman, N.~D., Prentice, R.,
  Lopez, A.~D., and et~al. (2013).
\newblock 50-year trends in smoking-related mortality in the {United States}.
\newblock {\em New England Journal of Medicine}, 368(4):351--364.

\bibitem[Tolley et~al., 1991]{Tolley:1991}
Tolley, H., Crane, L., and Shipley, N. (1991).
\newblock Strategies to control tobacco use in the united states -- a blueprint
  for public health action in the 1990s.
\newblock NIH publication no. 92- 3316 pp. 75 -- 144. Bethesda, Maryland: U.S.
  Department of Health and Human Services, Public Health Service, National
  Institutes of Health, National Cancer Institute.

\bibitem[van~der Vaart, 2000]{vanderVaart:2000book}
van~der Vaart, A. (2000).
\newblock {\em Asymptotic Statistics}.
\newblock Cambridge University Press.

\bibitem[VanderWeele and Ding, 2017]{vanderweele2017sensitivity}
VanderWeele, T.~J. and Ding, P. (2017).
\newblock Sensitivity analysis in observational research: introducing the
  e-value.
\newblock {\em Annals of internal medicine}, 167(4):268--274.

\bibitem[Vansteelandt et~al., 2008]{Vansteelandt:2008aa}
Vansteelandt, S., VanderWeele, T.~J., Tchetgen~Tchetgen, E.~J., and Robins,
  J.~M. (2008).
\newblock Multiply robust inference for statistical interactions.
\newblock {\em Journal of the American Statistical Association},
  103(484):1693--1704.

\bibitem[Wang and Zivot, 1998]{Wang:1998aa}
Wang, J. and Zivot, E. (1998).
\newblock Inference on structural parameters in instrumental variables
  regression with weak instruments.
\newblock {\em Econometrica}, 66(6):1389--1404.

\bibitem[Wang and Tchetgen~Tchetgen, 2018]{Wang:2018aa}
Wang, L. and Tchetgen~Tchetgen, E. (2018).
\newblock Bounded, efficient and multiply robust estimation of average
  treatment effects using instrumental variables.
\newblock {\em Journal of the Royal Statistical Society: Series B (Statistical
  Methodology)}, 80(3):531--550.

\bibitem[Warren et~al., 2014]{Warren:2014aa}
Warren, G.~W., Alberg, A.~J., Kraft, A.~S., and Cummings, K.~M. (2014).
\newblock The 2014 surgeon general's report:``the health consequences of
  smoking--50 years of progress": a paradigm shift in cancer care.
\newblock {\em Cancer}, 120(13):1914--1916.

\bibitem[Wooldridge, 2010]{wooldridge2010econometric}
Wooldridge, J.~M. (2010).
\newblock {\em Econometric analysis of cross section and panel data}.
\newblock MIT press.

\bibitem[Zhao et~al., 2019]{zhao:2019sens}
Zhao, Q., Small, D.~S., and Bhattacharya, B.~B. (2019).
\newblock Sensitivity analysis for inverse probability weighting estimators via
  the percentile bootstrap.
\newblock {\em Journal of the Royal Statistical Society: Series B (Statistical
  Methodology)}, 81(4):735--761.

\end{thebibliography}

\newpage

\setcounter{equation}{0}
\setcounter{table}{0}
\setcounter{lemma}{1}
\setcounter{section}{0}
\renewcommand{\theequation}{S\arabic{equation}}
\renewcommand{\thetheorem}{S\arabic{theorem}}
\renewcommand{\theequation}{S\arabic{equation}}
\renewcommand{\thetable}{S\arabic{table}}
\renewcommand{\theassumption}{S\arabic{assumption}}
\renewcommand{\thesection}{S\arabic{section}}
\renewcommand{\thefigure}{S\arabic{figure}}

\begin{center}
{\Large\bf Supplementary Materials}
\end{center}

\section{Additional Results for Instrumented DID}
\subsection{Instrumented DID when treatment effect may change over time}

If  Assumption \ref{assum: 1} and Assumption \ref{assump: iv for trend}(a)-(c) hold, then
\begin{align}
	\frac{\delta_Y(\X)}{\delta_D(\X)}=  \frac{E( D_1^{(1)}- D_1^{(0)}| \X)}{\delta_D(\X) } E(Y_1^{(1)}- Y_1^{(0)}|\X)- \frac{E( D_0^{(1)}- D_0^{(0)}|\X)}{\delta_D(\X)} E(Y_0^{(1)}- Y_0^{(0)}|\X).  \label{seq: treatment effect not stable}
\end{align} 
Consider the case without observed covariates. When the treatment effect  may vary over time, $ \delta_Y/\delta_D $ still has a nice interpretation under some special scenarios: (i) when either $  E( D_0^{(1)}- D_0^{(0)}) $ or  $ E( D_1^{(1)}- D_1^{(0)})$ is zero, then $ \delta_Y/\delta_D $ is the average treatment effect at the time point $ t $ in which $  E( D_t^{(1)}- D_t^{(0)}) \neq 0$; (ii) when $  E( D_0^{(1)}- D_0^{(0)}) $ and  $ E( D_1^{(1)}- D_1^{(0)})$  are both non-zero and of opposite sign, then $ E( D_0^{(1)}- D_0^{(0)})/\delta_D\in(-1,0)$ and   $ \delta_Y/\delta_D $ is a weighted average of $  E(Y_1^{(1)}- Y_1^{(0)})$ and $ E(Y_0^{(1)}- Y_0^{(0)}) $ with non-negative weights. Otherwise, although $ \delta_Y/\delta_D $ is still a weighted average of the treatment effects at the two time points,  
the weights can be negative and  $ \delta_Y/\delta_D $ no longer has a clear causal interpretation. For instance, if $ E(Y_1^{(1)}- Y_1^{(0)})> E(Y_0^{(1)}- Y_0^{(0)}) $ and $E( D_1^{(1)}- D_1^{(0)})> E( D_0^{(1)}- D_0^{(0)})>0 $, then $ 	\delta_Y/\delta_D>E(Y_1^{(1)}- Y_1^{(0)})> E(Y_0^{(1)}- Y_0^{(0)}) $, i.e., $ \delta_Y/\delta_D $  is larger than any time-specific average treatment effect. 

\subsection{One-sample and two-sample Wald estimators}

Let $ \xrightarrow{d} $ denote convergence in distribution. Theorem \ref{theo: Wald} establishes the asymptotic property for  the one-sample instrumented DID Wald estimator
$ \hat{\beta} $. 
\begin{theorem}\label{theo: Wald}
	Under Assumptions \ref{assum: 1} and \ref{assump: iv for trend},  and assume the second moments are finite, as $ n\rightarrow \infty $, the Wald estimator $ \hat{\beta}$ in   (\ref{eq: Wald estimator})  is consistent and asymptotically normal, i.e., 
	\begin{align}
		| \delta_D| \sqrt{n}(\hat{\beta}-\beta_{0}) \xrightarrow{d} N\left(0, \sum_{t= 0,1} \sum_{z= 0,1} \frac{\var(Y-\beta_0 D|T=t,Z=z)}{P(T=t, Z=z)}\right).\label{eq: CLT Wald}
	\end{align}
\end{theorem}

For statistical inference, we can use a consistent plug-in variance estimator 
\begin{align}
	\frac{1}{n(\hat{\delta}_{D})^2}\sum_{t=0,1} \sum_{z= 0,1} \frac{\widehat\var(Y-\hat\beta D|T=t,Z=z)}{\hat P(T=t, Z=z)}, \label{eq: Wald variance estimator}
\end{align} 
where $ \hat{\delta}_D$ is defined in (\ref{eq: Wald estimator}), $ \hat{P}(T=t, Z=z)= \sum_{i=1}^{n} I(T_i=t,  Z_i=z)/n$, $ \widehat{\var}(Y- \hat{\beta}D| T=t, Z=z) $ is the sample variance of $ Y_i- \hat{\beta}D_i $ within the stratum with $ T_i=t, Z_i=z $.

The next theorem establishes the asymptotic property for  the two-sample instrumented DID Wald estimator $	\hat{\beta}_{\rm TS} $. 

\begin{theorem}\label{theo: TSWald}
	Suppose that Assumptions \ref{assum: 1} and \ref{assump: iv for trend} hold for both $ (T_a, Z_a, D_a, Y_a) $ and  $ (T_b, Z_b, D_b, Y_b) $, and $ E(Y_{a}|T_{a}, Z_{a}) =E(Y_{b}|T_{b}, Z_{b}) $,  $E(D_{a}|T_{a}, Z_{a}) =E(D_{b}|T_{b}, Z_{b}) $. Also assume that $ \lim_{n_a, n_b\rightarrow\infty} \min(n_a, n_b)/n_c=\alpha_c\geq 0 $ for $ c\in\{a, b\} $,  and the second moments are finite.	As $ \min (n_a, n_b) \rightarrow \infty $, the two-sample Wald estimator $ \hat{\beta}_{\rm TS}$  is consistent and asymptotically normal, i.e., 
	\begin{align}
		&|\delta_{Db}|\sqrt{\min (n_a, n_b)}  (\hat{\beta}_{\rm TS}- \beta_0) \xrightarrow{d} \nonumber\\
		&\qquad N \left(0, \  \sum_{t=0,1}\sum_{z=0,1}\alpha_a \frac{\var(Y_a|T_a=t, Z_a=z)}{	 P(T_a=t, Z_a=z)} + \alpha_b \beta_0^2 \frac{\var(D_b|T_b=t, Z_b=z)}{P(T_b=t, Z_b=z)}\right). \label{seq: TS wald}
	\end{align}
\end{theorem}

For statistical inference, a consistent plug-in variance estimator for $ \hat{\beta}_{\rm TS} $ is 
\begin{eqnarray}
	\frac{1}{(\hat\delta_{Db})^2} \sum_{t=0,1}\sum_{z=0,1}\left[ \widehat \var\{\hat\mu_{Ya}(t, z)\}+ \hat\beta_{\rm TS}^2 \widehat \var\{\hat\mu_{Db}(t, z)\} \right], \label{seq: se for TS wald}
\end{eqnarray}
where $\hat\mu_{Ya}(t, z)$ and $\hat\mu_{Db}(t, z)$ are as defined in \eqref{eq: Wald estimator} but evaluated respectively at the outcome dataset and the exposure dataset,  $ \widehat\var\{\hat\mu_{Ya}(t, z)\}$ and $ \widehat\var\{\hat\mu_{Db}(t, z)\}$ are their consistent variance estimators. In fact, $ \hat\beta_{\rm TS} $ and its variance estimator can be calculated provided that these summary statistics are available. 

\subsection{Sensitivity analysis}
\label{subsec: sa}
We develop sensitivity analysis methods to evaluate how sensitive the conclusion is to violations of Assumption \ref{assump: iv for trend}(d) for both the one-sample and two-sample designs when there are no observed covariates. There is a large and growing literature on sensitivity analysis, e.g., \cite{Rosenbaum:1987aa, Imbens:2003, vanderweele2017sensitivity} and \cite{fogary2017avg}. 

Consider first the one-sample design. 
When  Assumption \ref{assump: iv for trend}(d) does not hold,  i.e., $  E( Y_1^{(1)}-Y_1^{(0)})\neq E(Y_0^{(1)}-Y_0^{(0)}) $. We use two sensitivity parameters $ \gamma_L, \gamma_U $ to quantify deviate from  Assumption \ref{assump: iv for trend}(d): $\Delta:= E( Y_1^{(1)}-Y_1^{(0)})- E(Y_0^{(1)}-Y_0^{(0)})\in [\gamma_L, \gamma_U]$, where $ \gamma_L\leq 0\leq \gamma_U $. When $ \gamma_L=\gamma_U $, it is the same as the case under Assumption \ref{assump: iv for trend}(d). Next, we construct a confidence interval for $  \beta^\ast = E(Y_0^{(1)}-Y_0^{(0)}) $ when $ \Delta\in [\gamma_L, \gamma_U] $; similar approach can be developed  for  $  E(Y_1^{(1)}-Y_1^{(0)}) $.

From \eqref{seq: treatment effect not stable}, we know that $ \beta^\ast  = \delta_Y/\delta_D - \Delta \{\mu_D(1,1) - \mu_D(1,0)\} /\delta_D $, whose sample analogue is defined as $ \hat  \beta_{\rm SA} (\Delta)=\hat  \delta_Y/\hat\delta_D - \Delta \{\hat\mu_D(1,1) - \hat\mu_D(1,0)\} /\hat \delta_D $. Similar to the proof of Theorem \ref{theo: Wald},  the asymptotic distribution of $ \hat  \beta_{\rm SA}(\Delta) $ is 
	\begin{align*}
	| \delta_D| \sqrt{n}( \hat  \beta_{\rm SA}(\Delta) -\beta^\ast ) \xrightarrow{d} N\left(0, \sum_{t= 0,1} \sum_{z= 0,1} \frac{\var(Y-(\beta^\ast + t\Delta) D|T=t,Z=z)}{P(T=t, Z=z)}\right).
\end{align*}
Denote a consistent variance estimator of $ \hat  \beta_{\rm SA}(\Delta) $  as $ \hat V_{\rm SA} (\Delta) $, let $ C_L(\Delta)= \hat  \beta_{\rm SA}(\Delta) - 1.96 \hat V_{\rm SA} (\Delta)^{1/2}  $ and  $ C_U(\Delta)= \hat  \beta_{\rm SA}(\Delta) + 1.96 \hat V_{\rm SA} (\Delta)^{1/2}  $, then $ [C_L(\Delta), C_U(\Delta)] $ is an asymptotic 95\% confidence interval for $\beta^\ast $ at any given value of $ \Delta $. In addition, by applying the union method \citep{zhao:2019sens}, we have that $ \big[ \inf_{\Delta \in [\gamma_L, \gamma_U]} C_L(\Delta) ,\sup_{\Delta \in [\gamma_L, \gamma_U]} C_U(\Delta)  \big]  $
is an asymptotic confidence interval with at least 95\% coverage for any $ \Delta\in  [\gamma_L, \gamma_U]$. 

The sensitivity analysis for the two-sample setting is analogous. Define $ \hat  \beta_{\rm SA,TS} (\Delta)=\hat  \delta_{Ya}/\hat\delta_{Db} - \Delta \{\hat\mu_{Db}(1,1) - \hat\mu_{Db}(1,0)\} /\hat \delta_{Db}$. Similar to the proof of Theorem \ref{theo: TSWald}, the asymptotic distribution of  $  \hat  \beta_{\rm SA,TS} (\Delta) $ is 
	\begin{align}
	&|\delta_{Db}|\sqrt{\min (n_a, n_b)}  (\hat  \beta_{\rm SA,TS} (\Delta) - \beta^*) \xrightarrow{d} \nonumber\\
	&\qquad N \left(0, \  \sum_{t=0,1}\sum_{z=0,1}\alpha_a \frac{\var(Y_a|T_a=t, Z_a=z)}{	 P(T_a=t, Z_a=z)} + \alpha_b (\beta^*+ t\Delta)^2 \frac{\var(D_b|T_b=t, Z_b=z)}{P(T_b=t, Z_b=z)}\right). \nonumber 
\end{align}
The construction of the confidence interval follows the same step as the one-sample design. 

%Following \cite{Rosenbaum:1987aa}, 
%it is common to report the ``tipping point'' -- the values of  $ (\gamma_L, \gamma_U) $ under which the confidence interval starts to cover zero and we fail to conclude with statistical significance.  

\section{Technical Proofs}
\label{sec: proofs}

\subsection{Proof of Proposition \ref{prop: iv for trend, Wald}}
\begin{proof}
	First, note that for $ z=0,1, $
	\begin{align}
	&	\mu_Y(1,z,X) -  	\mu_Y(0,z,X)  \nonumber\\
		& E(Y|X, T=1, Z=z) - E(Y|X, T=0, Z=z)  \nonumber\\
		&=E(Y_1^{(D_1^{(z)})} |X, T=1, Z=z) - E(Y_0^{(D_0^{(z)})} |X,  T=0, Z=z)  \nonumber\\
		&=E(Y_1^{(D_1^{(z)})} - Y_0^{(D_0^{(z)})}|X, Z=z)  \nonumber\\
		&= E(D_1^{(z)}Y_1^{(1)}+ (1-D_1^{(z)} ) Y_1^{(0)}- D_0^{(z)} Y_0^{(1)} - (1-D_0^{(z)} )  Y_0^{(0 )} |X,  Z=z) \nonumber\\ 
		&= E(D_1^{(z)} (Y_1^{(1)}-Y_1^{(0)}) - D_0^{(z)}  (Y_0^{(1)} - Y_0^{(0)}) +Y_1^{(0)}-  Y_0^{(0)} | X, Z=z) \nonumber\\
		&= E(D_1^{(z)} (Y_1^{(1)}-Y_1^{(0)}) - D_0^{(z)}  (Y_0^{(1)} - Y_0^{(0)})| X) + E(Y_1^{(0)}-  Y_0^{(0)}| X )\nonumber 
	\end{align}
where the second line is from Assumption \ref{assum: 1}(a), the third line is  from Assumption \ref{assum: 1}(c), the last line is from Assumption \ref{assump: iv for trend}(b). Thus, 
	\begin{align*}
		\delta_Y(X)&=  E((D_1^{(1)} - D_1^{(0)} ) (Y_1^{(1)}-Y_1^{(0)})|X)- E((D_0^{(1)} - D_0^{(0)} ) (Y_0^{(1)}-Y_0^{(0)})|X) \nonumber \\
		&=   E(D_1^{(1)} - D_1^{(0)} |X) E (Y_1^{(1)}-Y_1^{(0)}|X)- E(D_0^{(1)} - D_0^{(0)} |X) E (Y_0^{(1)}-Y_0^{(0)}|X)  \nonumber \\
		&=    E(D_1^{(1)} - D_1^{(0)} - D_0^{(1)}  + D_0^{(0)}|X) \beta_0(X) \\
		&= \delta_D(X)  \beta_0(X), 
	\end{align*}
where the second line is from Assumption  \ref{assump: iv for trend}(c), the third line is from Assumption  \ref{assump: iv for trend}(d), the last line again uses Assumption  \ref{assum: 1}(a)-(b).

\end{proof} 

\subsection{Derivation of $ \delta_Y(X)/\delta_D (X)$ under the monotonicity assumption}
From the proof of Proposition \ref{prop: iv for trend, Wald} and under the monotonicity assumption stated in the main article, we have 
\begin{align*}
	\delta_Y (X) &=  E( (D_1^{(1)}- D_1^{(0)} ) (Y_1^{(1)}- Y_1^{(0)} )|X)- E( (D_0^{(1)}- D_0^{(0)} ) (Y_0^{(1)}- Y_0^{(0)} )|X) \nonumber\\
	&=E( Y_1^{(1)}- Y_1^{(0)} \mid  D_1^{(1)}- D_1^{(0)}  = 1,X) P ( D_1^{(1)}- D_1^{(0)}  = 1 |X) \nonumber\\
	&\qquad - E( Y_0^{(1)}- Y_0^{(0)} \mid  D_0^{(1)}- D_0^{(0)} =1|X)  P( D_0^{(1)}- D_0^{(0)}  = 1|X) \nonumber\\
	&= E( Y_t^{(1)}- Y_t^{(0)} \mid  D_t^{(1)}- D_t^{(0)}  = 1)   \left\{ P ( D_1^{(1)}- D_1^{(0)}  = 1 |X) - P( D_0^{(1)}- D_0^{(0)}  = 1|X) \right\}\nonumber,
\end{align*}
where the last line is from the assumption that $ E( Y_1^{(1)}- Y_1^{(0)} \mid  D_1^{(1)}- D_1^{(0)}  = 1) = E( Y_0^{(1)}- Y_0^{(0)} \mid  D_0^{(1)}- D_0^{(0)}  = 1) $.  In addition, $ \delta_D(X) =P ( D_1^{(1)}- D_1^{(0)}  = 1|X ) - P( D_0^{(1)}- D_0^{(0)}  = 1|X)$.  This completes the proof.

\subsection{Proof of Theorem \ref{theo: Wald}}
From the definition of $ \hat\beta $, 
\begin{align}
	\sqrt{n}(\hat{\beta}-\beta_0)&=\frac{\sqrt{n} (\hat{\delta}_Y- \beta_0\hat \delta_D)}{\hat{\delta}_D} \nonumber.
\end{align}
Let $ \mathcal{F}= \{ T_i, Z_i, i=1,\dots, n\} $ and 
\begin{align}
	K_i&= \sqrt{n}(Y_i- \beta_0 D_i)\bigg\{  \frac{I(T_i=1, Z_i=1)}{\sum_{i=1}^{n}I(T_i=1, Z_i=1) }- \frac{I(T_i=1, Z_i=0)}{\sum_{i=1}^{n}I(T_i=1, Z_i=0) }\nonumber\\
	&\qquad\qquad \qquad\qquad- \frac{I(T_i=0, Z_i=1)}{\sum_{i=1}^{n}I(T_i=0, Z_i=1) }+ \frac{I(T_i=0, Z_i=0)}{\sum_{i=1}^{n}I(T_i=0, Z_i=0) } \bigg\} \nonumber. 
\end{align}
Then, we can write 
\begin{align}
	\sqrt{n}(\hat{\delta}_Y- \beta_0\hat \delta_D)= \sum_{i=1}^n K_i\nonumber.
\end{align}
First, note that $ K_i $, $i=1,\dots, n  $ are independent conditional on $  \mathcal{F}$, and $ E( \sum_{i=1}^{n}K_i|  \mathcal{F})=\sqrt{n}(  \delta_Y- \beta_0 \delta_D)=0$, and 
\begin{align}
	\var(K_i|\mathcal{F})=n \sum_{t=0}^{1} \sum_{z=0}^{1} \var(Y- \beta_0D|T=t, Z=z)\frac{I(T_i=t, Z_i=z)}{\{ \sum_{i=1}^{n}I(T_i=t, Z_i=z)\}^2} \nonumber.
\end{align}
We prove that $ \sum_{i=1}^{n} K_i$ is asymptotically normal by verifying Lindeberg's condition. Let 
\begin{align}
	\sigma^2= \sum_{i=1}^{n}\var(K_i|\mathcal{F})=  \sum_{t=0}^{1} \sum_{z=0}^{1} \frac{\var(Y- \beta_0D|T=t, Z=z)}{n^{-1}\sum_{i=1}^{n}I(T_i=t, Z_i=z) } \nonumber, 
\end{align}
we have that 
\begin{align}
	&\frac{\max_i \var(K_i|\mathcal{F})}{\sigma^2}=\max_{t', z'}\frac{ \frac{ \text{Var}(Y- \beta_0 D|T=t', Z=z')}{\{ \sum_{i=1}^{n} I(T_i=t', Z_i=z')\}^2} }{\sum_{z=0}^{1}\sum_{z=0}^{1} \frac{\text{Var}(Y- \beta_0D|T=t, Z=z)}{\sum_{i=1}^{n}I(T_i=t, Z_i=z) }  } \leq \max_{t', z'}\frac{ \frac{ \text{Var}(Y- \beta_0 D|T=t', Z=z')}{\{ \sum_{i=1}^{n} I(T_i=t', Z_i=z')\}^2} }{\frac{\text{Var}(Y- \beta_0D|T=t', Z=z')}{\sum_{i=1}^{n}I(T_i=t', Z_i=z') }  }  \nonumber \\
	&=  \max_{t', z'}\frac{1}{\sum_{i=1}^{n}I(T_i=t', Z_i=z') } =o(1). \nonumber 
\end{align}
Hence, for any $ \epsilon>0 $,
\begin{align}
	&\sum_{i=1}^{n} E\left\{ \frac{(K_i- E(K_i|\mathcal{F}))^2}{\sigma^2} I\left( |K_i - E(K_i |\mathcal{F})|> \epsilon\sigma \right)\mid\mathcal{F}\right\}\nonumber\\
	&=\sum_{i=1}^{n} \frac{\var(K_i|\mathcal{F})}{\sigma^2}E\left\{  \frac{(K_i- E(K_i|\mathcal{F}))^2}{\var(K_i|\mathcal{F})}I\left( |K_i - E(K_i |\mathcal{F})|> \epsilon\sigma \right)\mid\mathcal{F}\right\}\nonumber\\
	&\leq \max_i E\left\{  \frac{(K_i- E(K_i|\mathcal{F}))^2}{\var(K_i|\mathcal{F})}I\left( \frac{|K_i - E(K_i |\mathcal{F})|}{\sqrt{\var(K_i|\mathcal{F})}}> \frac{\epsilon\sigma}{\sqrt{\var(K_i|\mathcal{F})} } \right)\mid\mathcal{F}\right\}\nonumber\\
	&=o(1)\nonumber,
\end{align}
where the last equality is from dominated convergence theorem and the facts that $ \{K_i- E(K_i|\mathcal{F}) \}/ \sqrt{\var(K_i|\mathcal{F})}$ has expectation zero and variance 1 conditional on $ \mathcal{F} $, and \\$  \max_i \var(K_i|\mathcal{F})/\sigma^2=o(1)$. Therefore,  Lindeberg's condition holds.  Applying Linderberg Central Limit Theorem, we have  that conditional on $ \mathcal{F} $,
\[
\frac{\sqrt{n}(\hat{\delta}_Y- \beta_0\hat \delta_D)}{\sigma } \mid \mathcal{F}\xrightarrow{d} N(0,1).
\]
By a dominated convergence argument, we have that the above equation also  holds unconditionally. Then, by weak law of large numbers and Slutsky's theorem, it is easy to show that 
\[
\sigma^2= \sum_{t=0}^{1} \sum_{z=0}^{1} \frac{\var(Y- \beta_0D|T=t, Z=z)}{P (T=t, Z=z) }  +o_p(1),
\] 
and 
\[
\sqrt{n}(\hat{\delta}_Y- \beta_0\hat \delta_D)\xrightarrow{d} N\left(0,\sum_{t=0}^{1} \sum_{z=0}^{1} \frac{\var(Y- \beta_0D|T=t, Z=z)}{P (T=t, Z=z) }\right).
\] 
Finally, we can similarly show that $ \sqrt{n} (\hat{\delta}_D- \delta_D)$ is asymptotically normal, which implies that $ \hat{\delta}_D \xrightarrow{p} \delta_D$. Again using Slutsky's theorem, we have proved (\ref{eq: CLT Wald}).

\subsection{Proof of Theorem \ref{theo}}
In this section, we use subscripts to explicitly index quantities that depend on the distribution $ P $, we use a zero subscript to denote a quantity evaluated at the true distribution $ P=P_0 $, we use a $ \epsilon $ subscript to denote a quantity evaluated at the parametric submodel $ P=P_\epsilon $. We will show that $ \varphi ({O}; \bpsi_P, \bmeta_P) $ is proportional to the efficient influence function by showing that it is the canonical gradient of the pathwise derivative of $ \bpsi_P $, i.e, 
\begin{eqnarray}
	\frac{\partial \bpsi_\epsilon}{\partial \epsilon} \bigg|_{\epsilon=0}= C_0^{-1} E_0 \left\{\varphi ( {O}; \bpsi_P, \bmeta_P) s_0(\bmO) \right\},
\end{eqnarray}
where $ \bpsi_\epsilon= \bpsi_{P_\epsilon} $, $ s_\epsilon(\bmO)= \partial \log dP_\epsilon (\bmO) /\partial \epsilon $ denotes the parameter submodel score, $ C_0 $ is defined later in (\ref{eq: C_0}).

By definition, we have
\[
\bpsi_P= \arg\min_{\bpsi}  \int w(\bmv) \{ \beta_P(\bmv)- \beta(\bmv; \bpsi)\}^2 dP(\bmv),
\]
and thus
\[
\int q(\bmv; \bpsi) \{ \beta_P(\bmv)- \beta(\bmv; \bpsi)\} dP(\bmv)=0, 
\]
where $ q(\bmv; \bpsi) =w(\bmv) \frac{\partial \beta(\bmv; \bpsi)}{\partial \bpsi} $.
Evaluating the above at $ P=P_\epsilon $ gives 
\[
\int q(\bmv; \bpsi_\epsilon) \{ \beta_\epsilon(\bmv)- \beta(\bmv; \bpsi_\epsilon)\} dP_\epsilon(\bmv)=0,
\]
Differentiating the above with respect to $ \epsilon $ using the chain rule and evaluating at the truth $ \epsilon=0 $ give
\begin{align}
	&\int   \frac{\partial q(\bmv; \bpsi)}{\partial \bpsi} \bigg|_{\bpsi=\bpsi_0} \frac{\partial \bpsi_\epsilon}{\partial \epsilon}\bigg|_{\epsilon=0} \left\{ \beta_0(\bmv)- \beta(\bmv; \bpsi_0)\right\} dP_0(\bmv) \nonumber\\
	&+ \int q(\bmv; \bpsi_0)\left\{\frac{\partial \beta_\epsilon(\bmv)}{\partial \epsilon}\bigg|_{\epsilon=0} -  \frac{\partial \beta(\bmv; \bpsi)}{\partial \bpsi} \bigg|_{\bpsi=\bpsi_0} \frac{\partial \bpsi_\epsilon}{\partial \epsilon}\bigg|_{\epsilon=0} \right\} dP_0(\bmv) \nonumber\\
	&+\int  q(\bmv; \bpsi_0)\left\{ \beta_0(\bmv) - \beta(\bmv;\bpsi_0)\right\} s_0(\bmv) dP_0(\bmv)=0 \nonumber.
\end{align}
Rearranging the above equation, we have 
\begin{align}
	&\frac{\partial \bpsi_\epsilon}{\partial \epsilon}\bigg|_{\epsilon=0}\underbrace{\int\left[ \frac{\partial q(\bmv; \bpsi)}{\partial \bpsi} \bigg|_{\bpsi=\bpsi_0}\left\{ \beta_0(\bmv)- \beta(\bmv; \bpsi_0) \right\}   - q(\bmv; \bpsi_0) \frac{\partial \beta(\bmv; \bpsi)}{\partial \bpsi}\bigg|_{\bpsi=\bpsi_0}\right]dP_0(\bmv)}_{-C_0}\label{eq: C_0} \\ 
	&+ \int q(\bmv; \bpsi_0)\left\{\frac{\partial \beta_\epsilon(\bmv)}{\partial \epsilon}\bigg|_{\epsilon=0}+\left\{ \beta_0(\bmv) - \beta(\bmv;\bpsi_0)\right\} s_0(\bmv)  \right\} dP_0(\bmv) \nonumber=0,
\end{align}
and thus
\begin{align}
	&C_0\frac{\partial \bpsi_\epsilon}{\partial \epsilon}\bigg|_{\epsilon=0}=  \int q(\bmv; \bpsi_0)\left\{\frac{\partial \beta_\epsilon(\bmv)}{\partial \epsilon}\bigg|_{\epsilon=0}+\left\{ \beta_0(\bmv) - \beta(\bmv;\bpsi_0)\right\} s_0(\bmv)  \right\} dP_0(\bmv) \nonumber.
\end{align}

Next, we will derive $ \frac{\partial \beta_\epsilon(\bmv)}{\partial \epsilon}|_{\epsilon=0} $. Note that
\begin{align}
	&\frac{\partial \beta_\epsilon(\bmv)}{\partial \epsilon}\bigg|_{\epsilon=0} \nonumber\\
	& = \frac{\partial }{\partial \epsilon} E_\epsilon\left[\frac{\delta_{Y\epsilon}(\X)}{\delta_{D\epsilon}(\X)}\bigg| \bmV=\bmv\right] \bigg|_{\epsilon=0} \nonumber\\
	&=\frac{\partial}{\partial \epsilon} \int \frac{\delta_{Y\epsilon}(\X)}{\delta_{D\epsilon}(\X)} dP_\epsilon(\X|\bmV=\bmv) \bigg|_{\epsilon=0}\nonumber\\
	&=\int \left[ \frac{\frac{\partial \delta_{Y\epsilon}(\X) }{\partial\epsilon} |_{\epsilon=0} \delta_{D0}(\X)- \delta_{Y0}(\X)  \frac{\partial \delta_{D\epsilon}(\X) }{\partial\epsilon} |_{\epsilon=0}}{[\delta_{D0}(\X)]^2}+\frac{\delta_{Y0}(\X)}{\delta_{D0}(\X)} s_0(\X|\bmV) \right]dP_0(\X|\bmV=\bmv)  \nonumber, 
\end{align}
and 
\begin{align}
	&\frac{\partial \delta_{Y\epsilon}(\X) }{\partial\epsilon} \bigg|_{\epsilon=0} \nonumber\\
	&= E_0[ Ys_0 (Y|T,Z,\X) | T=1, Z=1, \X] - E_0[ Ys_0 (Y|T,Z,\X) | T=0, Z=1, \X] \nonumber\\
	&\qquad- E_0[ Ys_0 (Y|T,Z,\X) | T=1, Z=0, \X] +E_0[ Ys_0 (Y|T,Z,\X) | T=0, Z=0, \X] \nonumber\\
	& =E_0\left[\left\{ \frac{T Z}{P_0(T=1, Z=1|\X)} - \frac{(1-T) Z}{P_0(T=0, Z=1|\X)} \right.\right.\nonumber\\
	&\qquad\qquad\left.\left.- \frac{T (1-Z)}{P_0(T=1, Z=0|\X)} + \frac{(1-T) (1-Z)}{P_0(T=0, Z=0|\X)} \right\} Ys_0(Y|T,Z,\X) \bigg|\X\right] \nonumber \\
	&=E_0\left[ \frac{ (2Z-1) (2T-1)}{ \pi_0(T, Z, \X)} Ys_0(Y|T,Z,\X)|\X\right]  \nonumber,
\end{align}
where $ \pi_0(t, z, \X) = P_0(T=t, Z=z|\X)$. Similarly, we can also derive that 
\begin{align}
	&\frac{\partial \delta_{D\epsilon}(X) }{\partial\epsilon} \bigg|_{\epsilon=0} =E_0\left[ \frac{ (2Z-1) (2T-1)}{ \pi_0(T, Z, \X)}   Ds_0(D|T,Z,\X)|\X\right]  \nonumber.
\end{align}
Combining the above derivations, we have 
\begin{align}
	&C_0\frac{\partial \bpsi_\epsilon}{\partial \epsilon}\bigg|_{\epsilon=0}\nonumber\\
	&=  \int q(\bmv; \bpsi_0)\frac{\partial \beta_\epsilon(\bmv)}{\partial \epsilon}\bigg|_{\epsilon=0}dP_0(\bmv) +\int q(\bmv; \bpsi_0)\left\{ \beta_0(\bmv) - \beta(\bmv;\bpsi_0)\right\} s_0(\bmv) dP_0(\bmv)  \nonumber \\
	&=   \int q(\bmv; \bpsi_0) \bigg[ \frac{\frac{\partial \delta_{Y\epsilon}( \X) }{\partial\epsilon} |_{\epsilon=0} \delta_{D0}( \X)- \delta_{Y0}( \X)  \frac{\partial \delta_{D\epsilon}( \X) }{\partial\epsilon} |_{\epsilon=0}}{[\delta_{D0}( \X)]^2} \nonumber\\
	& \qquad \qquad\qquad\qquad\qquad\qquad\qquad\qquad+ \frac{\delta_{Y0}( \X)}{\delta_{D0}( \X)} s_0( \X|\bmV) \bigg]dP_0(\X|\bmV=\bmv)  dP_0(\bmv) \nonumber\\
	&\qquad+\int q(\bmv; \bpsi_0)\left\{ \beta_0(\bmv) - \beta(\bmv;\bpsi_0)\right\} s_0(\bmv) dP_0(\bmv).
\end{align}

We now turn to $ E_0 \left\{\varphi ( {O}; \bpsi_P, \bmeta_P) s_0(\bmO) \right\} $. Note that $ s_0(\bmO) $ is the parametric submodel score can be decomposed as
\[
s_0(\bmO)=s_0(Y,D|T, Z, \X) +s_0(T, Z|\X)+ s_0(\X|\bmV)+ s_0(\bmV).
\]
With the scaling factor, the efficient influence function is $ C_0^{-1} \varphi(\bmO; \bpsi_P, \bmeta_P) $,  where $ \varphi(\bmO; \bpsi_P, \bmeta_P) $ is defined in Theorem \ref{theo}. Therefore, 
\begin{align}
	&E_0\{\varphi(\bmO; \bpsi_0, \bmeta_0) s_0(\bmO)\} \nonumber\\
	&= E_0 \left\{ q(\bmV; \bpsi_0) \left[\frac{\delta_{Y0}(\X)}{\delta_{D0}(\X)}- \beta(\bmV; \bpsi_0)\right] \{s_0(\X|\bmV)+s_0({V})\}\right\} \nonumber\\
	&\qquad + E_0\left\{ q(\bmV; \bpsi_0) \frac{(2Z-1)(2T-1) }{\pi_0(T, Z,\X)\delta_{D0}(\X)} \left[ Y- E_0(Y|T,Z,\X) \right] s_0(Y|T,Z,\X)\right\}  \nonumber\\
	& \qquad - E_0\left\{ q(\bmV; \bpsi_0)  \frac{(2Z-1)(2T-1) }{\pi_0(T, Z,\X)\delta_{D0}(\X)}   \frac{\delta_{Y0}(\X)}{\delta_{D0}(\X)} [ D- E_0(D|T,Z,\X)]s_0(D|T, Z, \X)\right\} \nonumber\\
	&= E_0 \left\{ q(\bmV; \bpsi_0) \frac{\delta_{Y0}(\X)}{\delta_{D0}(\X)} 
	s_0(\X|\bmV) \right\} +E_0 \left\{ q(\bmV; \bpsi_0) \left[\beta_0(\bmV)- \beta(\bmV; \bpsi_0) \right]s_0({V})\right\} \nonumber\\
	&\qquad + E_0\left\{ q(\bmV; \bpsi_0)  \frac{(2Z-1)(2T-1) }{\pi_0(T, Z,\X)\delta_{D0}(\X)}  Ys_0(Y|T,Z,\X)\right\}  \nonumber\\
	& \qquad - E_0\left\{ q(\bmV; \bpsi_0)  \frac{(2Z-1)(2T-1) }{\pi_0(T, Z,\X)\delta_{D0}(\X)}  \frac{\delta_{Y0}(\X)}{\delta_{D0}(\X)}  Ds_0(D|T, Z, \X)\right\} \nonumber\\ 
	&= C_0 \frac{\partial \bpsi_\epsilon}{\partial \epsilon} \bigg|_{\epsilon=0} \nonumber,
\end{align}
where the derivations follow from $ E_0(s_0(\bmO_1| \bmO_2)| \bmO_2)=0$ for any $(\bmO_1, \bmO_2)\subset \bmO$ and iterated expectation. Hence, $ C_0^{-1} \varphi(\bmO; \bpsi_P, \bmeta_P) $ is the efficient influence function.

\subsection{Proof of multiple robustness }
From the definition of $ \bpsi_0 $ in (\ref{eqn: obj}),  it is true that 
\begin{eqnarray}
	E\left[ q(\bmV; \bpsi_0) \left\{ \beta_0(\bmV) - \beta(\bmV; \bpsi_0)\right\}\right] =0. \label{eq: estimation equation}
\end{eqnarray}

Under $ \calM_1 $, $ \bar \pi(T, Z, \X)= \pi_0(T, Z, \X) $, $ \bar\mu_D(T, Z, \X)= \mu_{D0}(T, Z, \X) $ and thus $ \bar{\delta}_D(\X) =\delta_{D0}(\X)$. Then,
\begin{align}
	&E [ \varphi( \bmO; \bpsi_0, \bar\bmeta)] \nonumber\\
	&= E\left[ q(\bmV; \bpsi_0) \left\{ \frac{\bar\delta_Y(\X)}{\delta_{D0} (\X)} - \beta(\bmV; \bpsi_0)\right\} \right]\nonumber\\
	&\qquad + E\left[  q(\bmV; \bpsi_0)\frac{(2Z-1)(2T-1) }{\pi_0(T, Z,\X)\delta_{D0}(\X)}  (Y-\bar\mu_Y (T, Z, \X)) \right]  \nonumber\\
	&\qquad - E\left[ q(\bmV; \bpsi_0)\frac{(2Z-1)(2T-1) }{\pi_0(T, Z,\X)\delta_{D0}(\X)}  \frac{\bar\delta_Y(\X)}{\delta_{D0}(\X)} [ D- \mu_{D0}(T, Z, \X)]\right] \nonumber\\
	&= E\left[ q(\bmV; \bpsi_0) \left\{ \frac{\bar\delta_Y(\X)}{\delta_{D0} (\X)} - \beta(\bmV; \bpsi_0)\right\} \right]+ E\left[ q(\bmV; \bpsi_0) \left\{ \frac{ \delta_{Y0}(\X)}{\delta_{D0}(\X)} -\frac{\bar \delta_{Y}(\X)}{\delta_{D0}(\X)}\right\} \right] \nonumber\\
	&= E\left[q(\bmV; \bpsi_0)\left\{  \beta_0(\bmV)- \beta(\bmV; \bpsi_0)\right\}  \right]=0, \nonumber
\end{align}
where the second equality uses the facts that $ E\{ C(2Z-1)(2T-1)/ \pi_0(T, Z, \X)  |\X\}=E\{ \mu_{C0}(T, Z, \X)(2Z-1)(2T-1)/ \pi_0(T, Z, \X)|\X  \}= \delta_{R0}(\X)$ and $ E\{\bar \mu_C(T, Z, \X)(2Z-1)(2T-1)/ \pi_0(T, Z, \X) |\X \}= \bar\delta_C(\X)$ for $ C\in\{ Y, D\} $.
Hence, the efficient influence function $ \varphi( \bmO; \bpsi, \bmeta) $ has expectation zero at $ \bpsi=\bpsi_0 $ under $ \calM_1 $.

Under $ \calM_2 $, $ \bar\pi(T, Z,\X)= \pi_0(T, Z, \X) $, $ \bar\delta_Y(\X)/ \bar\delta_D(\X) =\beta_0(\X)$. Then,
\begin{align}
	&E [ \varphi( \bmO; \bpsi_0,\bar \bmeta)]\nonumber\\
	&= E\left[ q(\bmV; \bpsi_0) \left\{ \beta_0(\X)- \beta(\bmV; \bpsi_0)\right\} \right] \nonumber\\
	&\qquad + E\left[  q(\bmV; \bpsi_0) \frac{(2Z-1)(2T-1)}{\pi_0 (T, Z, \X)\bar\delta_D(\X)} (Y- \bar\mu_Y(T, Z, \X))  \right]  \nonumber\\
	&\qquad - E\left[ q(\bmV; \bpsi_0) \frac{(2Z-1)(2T-1)}{\pi_0 (T, Z, \X)\bar\delta_D(\X)}  \beta_0(\X)[ D- \bar\mu_D(T, Z, \X)]\right] \nonumber\\
	&= E\left[ q(\bmV; \bpsi_0) \left\{ \beta_0(\X)- \beta(\bmV; \bpsi_0)\right\} \right]\nonumber\\
	& \qquad + E\left[ q(\bmV; \bpsi_0) \left\{ \frac{ \delta_{Y0}(\X)-\bar\delta_Y(\X) - \beta_0(\X) \delta_{D0}(\X) + \beta_0(\X)\bar \delta_D(\X)}{\bar\delta_D(\X)} \right\} \right] \nonumber\\
	&= E\left[q(\bmV; \bpsi_0)\left\{  \beta_0(\bmV)- \beta(\bmV; \bpsi_0)\right\}  \right] =0\nonumber.
\end{align}
Hence, the efficient influence function $ \varphi( \bmO; \bpsi, \bmeta) $ has expectation zero at $ \bpsi=\bpsi_0 $ under $ \calM_2 $.

Under $ \calM_3 $, $ \bar \mu_Y(T, Z, \X)=\mu_{Y0}(T, Z, \X)$, $ \bar \mu_D(T, Z, \X)=\mu_{D0}(T, Z, \X)$, and thus $ \bar\delta_Y(\X)/\bar\delta_D(\X)= \beta_0(\X) $. Then, 
\begin{align}
	E [ \varphi( \bmO; \bpsi_0,\bar \bmeta)]&= E\left[ q(\bmV; \bpsi_0) \left\{ \beta_0(\X)- \beta(\bmV; \bpsi_0)\right\} \right]+0=0 \nonumber,
\end{align}
where the first equality is from iterated expectations. Hence, the efficient influence function $ \varphi( \bmO; \bpsi, \bmeta) $ has expectation zero at $ \bpsi=\bpsi_0 $ under $ \calM_3 $.

\subsection{Proof of Theorem \ref{theo: 2}} 
In what follows, we will use $ P\{ f(\bmO)\} = \int f(\bmO) dP$ to denote expectation treating the function $ f $ as fixed; thus $ P\{ f(\bmO)\} $ is random when $ f $ is random, and is different from the fixed quantity $ E\{ f(\bmO)\} $ which averages over randomness in both $ f $ and $ \bmO $. 

Since $ \hat{\bpsi} $ is a $ Z $-estimator, using Theorem 5.31 of \cite{vanderVaart:2000book}, we have that under Assumption \ref{assump: semi}, 
\begin{align}
	&\sqrt{n} (\hat{\bpsi} - \bpsi_0) = - M_{\bpsi_0, \bar{\bmeta}}^{-1} \sqrt{n} P \{\varphi(\bmO; \bpsi_0, \hat{\bmeta})\} - M_{\bpsi_0, \bar{\bmeta}}^{-1} n^{-1/2} \sum_{i=1}^n\left[ \varphi(\bmO_i;\bpsi_0, \bar{\bmeta} ) - E\{ \varphi(\bmO;\bpsi_0, \bar{\bmeta} ) \} \right] \nonumber\\
	&\qquad \qquad \qquad +o_p(1+\sqrt{n} \| P \{\varphi(\bmO; \bpsi_0, \hat{\bmeta})\}\|) \nonumber,
\end{align}
where $ \|\beta\|= (\beta^T\beta)^{1/2} $ denotes the Euclidean norm. Using standard central limit theorem,  the second term is asymptotically normal, and is $ O_p(1) $. Hence, the consistency and rate of convergence of $ \hat{\bpsi}  $ depends on the property of the first term. We analyze $\sqrt{n} P \{\varphi(\bmO; \bpsi_0, \hat{\bmeta})\} $ in the following. 

For ease of exposition, we will simplify the notations to $q,  \mu_Y, \mu_D, \delta_Y, \delta_D, \pi $ and keep the involved random variables implicit.  Note that 
\begin{align}
	&P \{\varphi(\bmO; \bpsi_0, \hat{\bmeta})\}  \nonumber\\
	&=  P\left[q\left\{\frac{\hat{\delta}_Y}{ \hat{\delta}_D}-\beta_0(\X)+\frac{(2Z-1)(2T-1)}{\hat \pi  \hat{\delta}_D} \{ \mu_{Y0}- \hat{\mu}_Y - \frac{\hat{\delta}_Y}{\hat{\delta}_D} ( \mu_{D0}- \hat\mu_D )\}\right\}\right] \nonumber\\
	&=  P\left[\frac{q}{\hat{\delta}_D}\left\{ \hat{\delta}_Y-\beta_0(\X)\hat{\delta}_D+\frac{(2Z-1)(2T-1)}{\hat \pi  } \{ \mu_{Y0}- \hat{\mu}_Y - \frac{\hat{\delta}_Y}{\hat{\delta}_D} ( \mu_{D0}- \hat\mu_D )\}\right\}\right] \nonumber\\
	&=  P\left[\frac{q}{\hat{\delta}_D}\left\{ \hat{\delta}_Y-\beta_0(\X)\hat{\delta}_D  \right\}\right. \nonumber\\
	&\qquad ~ +\left. \frac{q}{\hat{\delta}_D} \left\{ \delta_{Y0}-\beta_0(\X)\delta_{D0}+\frac{(2Z-1)(2T-1)}{\hat \pi  } \{ \mu_{Y0}- \hat{\mu}_Y - \frac{\hat{\delta}_Y}{\hat{\delta}_D} ( \mu_{D0}- \hat\mu_D )\}\right\}\right] \nonumber\\
	&=  P\left[\frac{q}{\hat{\delta}_D}\left\{ \frac{(2Z-1)(2T-1)}{\pi_0} \{ \hat{\mu}_Y- \mu_{Y0}- \beta_0(\X) (\hat\mu_D- \mu_{D0})\} \right\}\right.\nonumber\\
	&\qquad ~ +\left.\frac{q}{\hat{\delta}_D}\left\{  \frac{(2Z-1)(2T-1)}{\hat \pi  } \{ \mu_{Y0}- \hat{\mu}_Y - \frac{\hat{\delta}_Y}{\hat{\delta}_D} ( \mu_{D0}- \hat\mu_D )\}\right\}\right] \nonumber\\
	&=  P\bigg[\frac{q (2Z-1)(2T-1)}{\hat{\delta}_D}\bigg\{ \frac{1}{\pi_0} \{ \hat{\mu}_Y- \mu_{Y0}- \beta_0(\X) (\hat\mu_D- \mu_{D0})\}  \nonumber\\
	& \qquad\qquad\qquad\qquad\qquad\qquad\qquad	+\frac{1}{\hat \pi  } \{ \mu_{Y0}- \hat{\mu}_Y - \frac{\hat{\delta}_Y}{\hat{\delta}_D} ( \mu_{D0}- \hat\mu_D )\}\bigg\}\bigg] \nonumber\\
	&=  P\bigg[\frac{q (2Z-1)(2T-1)}{\hat{\delta}_D}\bigg\{ (\frac{1}{\pi_0}- \frac{1}{\hat{\pi}}) \{ \hat{\mu}_Y- \mu_{Y0}- \beta_0(\X) (\hat\mu_D- \mu_{D0})\} \nonumber\\
	& \qquad\qquad\qquad\qquad\qquad\qquad\qquad +\frac{1}{\hat \pi  } \left(\beta_0(\X)- \frac{\hat\delta_Y}{\hat{\delta}_D}\right) (\mu_{D0}- \hat{\mu}_D)\bigg\}\bigg] \nonumber\\
	&=O_p\left(\| \hat{\pi} - \pi_0\|_2 \| \hat{\mu}_Y- \mu_{Y0} - \beta_0(\X) (\hat{\mu}_D- \mu_{D0})\|_2 +  \bigg\|\beta_0(\X)- \frac{\hat\delta_Y}{\hat{\delta}_D}\bigg\|_2 \| \mu_{D0}- \hat{\mu}_D\|_2\right)\nonumber\\
	&=O_p\left( \| \hat{\pi} - \pi_0\|_2 \left(\| \hat{\mu}_Y- \mu_{Y0}\|_2 + \|  (\hat{\mu}_D- \mu_{D0})\|_2 \right) +  \bigg\|\beta_0(\X)- \frac{\hat\delta_Y}{\hat{\delta}_D}\bigg\|_2 \| \mu_{D0}- \hat{\mu}_D\|_2 \right)\nonumber,
\end{align}
where the first equality is from  (\ref{eq: estimation equation}) and iterated expectation, the third equality is because $ \delta_{Y0}=\beta_0(\X)\delta_{D0}  $, the fourth equality is from the facts that $ P[ (2Z-1)(2T-1) \mu_{C0}/ \pi_0 |\X] = \delta_{C0}$ and $ P[ (2Z-1)(2T-1)\hat \mu_{C}/ \pi_0|\X] = \hat{\delta}_C$ for $ C\in\{ Y, D\} $, the second to the last line is from the	 Cauchy-Schwartz inequality that $ P(XY)\leq \|X\|_2 \|Y\|_2 $, the boundedness of $ q(\bmV; \bpsi_0) $, $ 1/ \hat{\delta}_D $ and $ 1/(\hat{\pi}\pi_0) $ (from the trend relevance assumption, the positivity assumption, and the Donsker condition), and  the fact that $ (2Z-1)^2(2T-1)^2=1 $, the last line is from the triangle inequality and the boundedness of $ \beta_0(\X) $. \\

\subsection{Proof of Theorem \ref{theo: TSWald}}
In this section, denote $ n_{\min}= \min\{n_a, n_b\} $.	From the definition of $ \hat{\beta}_{\rm TS} $, we have 
\begin{align}
	\sqrt{ n_{\min}}(\hat{\beta}_{\rm TS}-\beta_0)&=\frac{\sqrt{ n_{\min}} (\hat{\delta}_{Ya}- \beta_0\hat \delta_{Db})}{\hat{\delta}_{Db}} \nonumber. 
\end{align}
From the two-sample design, $ \hat\delta_{Ya}  $ is independent of $ \hat\delta_{Db}  $. Then, similar to the proof of Theorem 2, we can show that 
\[
\sqrt{n_a}( \hat \delta_{Ya} - \delta_{Ya} ) \xrightarrow{d} N\left(0, \  \sum_{t=0}^1 \sum_{z=0}^1 \frac{{\rm Var} (Y_a \mid T_a= t, Z_a=z )  }{P(T_a=t, Z_a=z)} \right),
\]
\[
\sqrt{n_b}( \hat \delta_{Db} - \delta_{Db} ) \xrightarrow{d} N\left(0, \  \sum_{t=0}^1 \sum_{z=0}^1 \frac{{\rm Var} (D_b \mid T_b= t, Z_b=z )  }{P(T_b=t, Z_b=z)} \right).
\]
In consequence, 
\begin{align*}
	\sqrt{ n_{\min}} \big\{ (\hat \delta_{Ya} -& \beta_0 \hat \delta_{Db} )- (\delta_{Ya}- \beta_0 \delta_{Db} )\big\} \xrightarrow{d} \\
	& N\left(0, \  \sum_{t=0}^1 \sum_{z=0}^1  \alpha_a \frac{{\rm Var} (Y_a \mid T_a= t, Z_a=z )  }{P(T_a=t, Z_a=z)}  +  \alpha_b \beta_0^2 \frac{{\rm Var} (D_b \mid T_b= t, Z_b=z )  }{P(T_b=t, Z_b=z)}  \right).
\end{align*}
Theorem \ref{theo: TSWald} follows from $ \delta_{Ya}- \beta_0 \delta_{Db} =\delta_{Ya}- \beta_0 \delta_{Da} =0   $, $ \hat\delta_{Db} = \delta_{Db}+ o_p(1)$ and Slutsky's theorem.

\section{Application}
\textsf{R} codes for constructing the dataset and reproducing the results are in  \textsf{smoking-lung.R} included in the supplementary materials. In the following, we provide additional details on the application.

\subsection{Data}
The 1970 NHIS data (\emph{personsx.rds}) were drawn using the \textsf{R lodown} package \\at {\tt http://asdfree.com}. The CDC mortality data were obtained from the CDC compressed mortality file. The mortality data are also included in the supplementary materials as \textsf{Compressed Mortality, 1975.txt, Compressed Mortality, 1985.txt, Compressed Mortality, 1995.txt, Compressed Mortality, 2005.txt}.

Standard errors for the cigarette smoking prevalence are obtained from the \textsf{survey} package in \textsf{R} to account for the NHIS complex sample design, following the variance estimation procedure available at {\tt https://www.cdc.gov/nchs/data/nhis/6372var.pdf} and also included in the supplementary materials as \textsf{6372var.pdf}. Standard errors for the lung cancer mortality rates are  calculated following
{\tt https://wonder.cdc.gov/wonder/help/cmf.html\#Standard-Errors}, using  the formula $\sqrt{ p/n}$, where $ p $ is the crude mortality rate, $ n $ is the sample size for the population. In Table \ref{tb: sample size}, we include the sample size for each birth cohort in each dataset. 
According to Theorem \ref{theo: TSWald} and Equation \eqref{seq: treatment effect not stable}, these obtained  standard errors suffice for constructing the consistent variance estimator for $ \hat\beta_{\rm TS} $.

\begin{table}
	\caption{Sample sizes for 1970 NHIS datasets and 1975, 1985, 1995, 2005 CDC WONDER compressed mortality datasets by birth cohort and gender \label{tb: sample size}} 
	\centering
	\begin{tabular}{lrrrr}
		\hline \\ [-1.5ex]
		Birth Cohorts & 1911-1920 & 1921-1930 & 1931-1940 & 1941-1950 \\ [0.5ex]\hline \\ [-1.5ex]
		\multicolumn{4}{l}{NHIS} \\[0.5ex]
		Men & 4,830 & 5,620 & 5,343 & 6,942 \\ 
		Women & 6,043 & 7,024 & 6,672 & 8,567 \\ 
		&&&&\\
		\multicolumn{4}{l}{CDC WONDER} \\[0.5ex]
		Men & 9,416,000 & 10,383,963 & 10,158,673 & 14,773,087 \\ 
		Women & 10,629,000 & 11,751,158 & 11,161,349 & 15,868,410 \\ 
		\hline
	\end{tabular}
\end{table}

\subsection{Use of gender as a surrogate for encouragement }
It is known that a standard IV does not need to have a causal effect on the exposure \citep{Hernan:2006aa}. It is also the case for the IV for DID; the IV for DID $ Z $ does not need to have a causal effect on the exposure; it suffices that the IV for DID is associated with the trend in exposure. 

Let $ D_t $ be the potential exposure that would be observed at time $ t $ if $ Z $ takes the value that naturally occurs. Using $ Z $ as a surrogate,  we can still establish the identification result in Proposition  \ref{prop: iv for trend, Wald} under Assumptions S1-S2 stated as follows. 

\begin{assumption}
	(a) (consistency) $ D= D_T $ and $ Y= Y_T^{(D)} $. \\
	(b) (positivity) $ 0< P(T=t, Z=z|X) <1 $ for $ t=0,1 $, $ z=0,1 $ with probability 1.\\
	(c) (random sampling) $ T\perp (D_t, Y_t^{(d)}, t=0,1, d=0,1) \mid Z,X$.
\end{assumption}
\begin{assumption}[Instrumented DID] With probability 1,  \\
	(a) (trend relevance) $ \delta_D\neq 0$.\\
	(b) (Independence \& exclusion restriction) $ Z\perp (Y_1^{(0)}  - Y_0^{(0)}, Y_t^{(1)}- Y_t^{(0)} , t=0,1)   |X$.  \\
	(c) (No unmeasured common effect modifier) $ E(D_t (Y_t^{(1)} - Y_t^{(0)})|X, Z=1)  - E(D_t (Y_t^{(1)} - Y_t^{(0)})|X, Z=0)  = (E(D_t|X, Z=1) -E(D_t|X, Z=0)  ) E(Y_t^{(1)} - Y_t^{(0)}) $ for $ t=0,1$.   \\
	(d) $ E(Y_1^{(1)} - Y_1^{(0)} |X)= E(Y_0^{(1)} - Y_0^{(0)} |X)$. 
\end{assumption}
Note that Assumption S2(c) is implied by Assumption \ref{assump: iv for trend}. To better understand Assumption S2(c), similar to \cite{Wang:2018aa},  assume in this paragraph only the existence of an unmeasured confounder $ U_t $ such that $ (D_t, Z) \perp (Y_t^{(1) } - Y_t^{(0)}) \mid U_t, X$ and $ Z\perp U_t|X $. Then, the same as the discussion of Theorem  \ref{assump: iv for trend}(c) in the main article,    Assumption S2(c) holds if either (i) there is no additive $ U_t $-$ Z$ interaction in $ E(D_t| Z, U_t, \X) $: $ E(D_t| Z=1, U_t, \X)- E(D_t| Z=0, U_t, \X) = E(D_t| Z=1, \X)- E(D_t| Z=0, \X) $; or (ii) there is no additive $ U_t$-$d $ interaction in $ E(Y^{(d)}| U_t, \X) $: $ E(Y^{(1)} - Y^{(0)}| U_t, \X) =  E(Y^{(1)} - Y^{(0)}| \X) $.

\subsection{Sensitivity analysis}
As mentioned in the main article, there is still concern about violating the stable treatment effect over time Assumption (Assumption \ref{assump: iv for trend}(d)), possibly because the cigarette design and composition have undergone changes that promote  deeper inhalation of smoke \citep{Thun:2013aa, Warren:2014aa}. In this section, we apply the  sensitivity analysis developed in Section \ref{subsec: sa}. 

\begin{figure}
	\centering
	\begin{subfigure}[b]{0.55\textwidth}
		\centering
		\includegraphics[width=\textwidth]{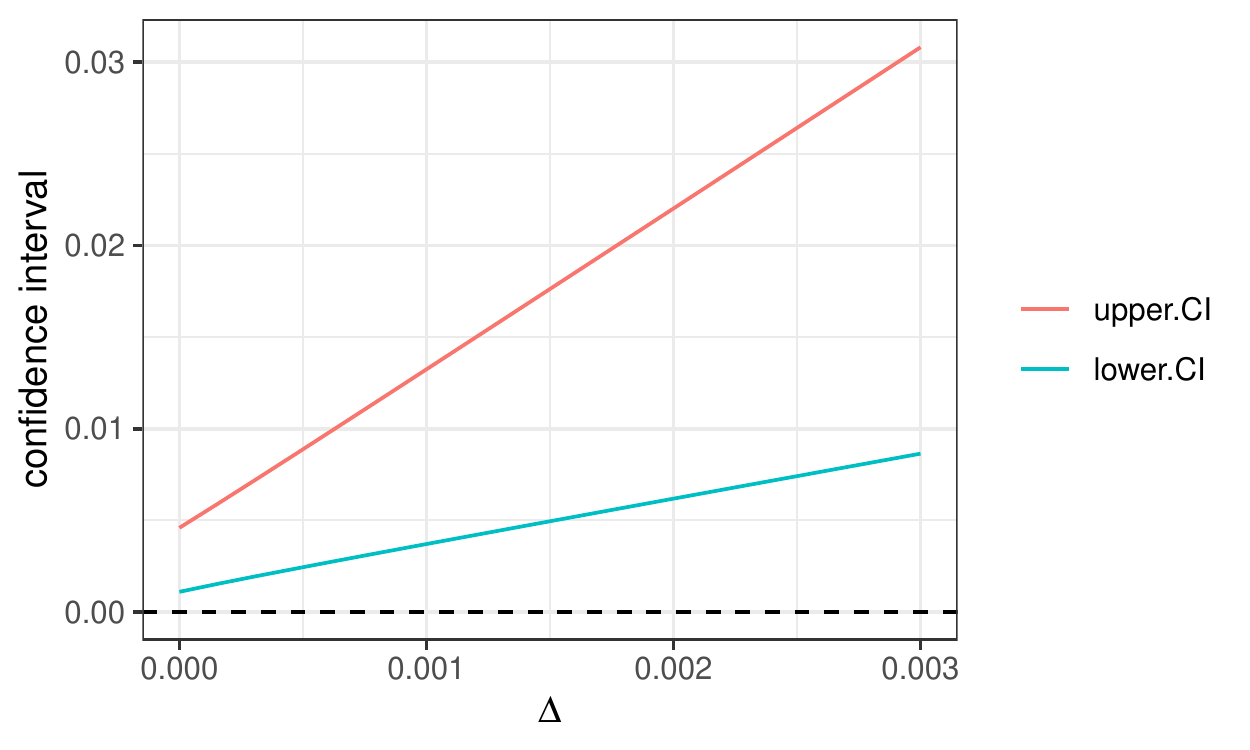}
		\caption{Birth cohorts: 1911-1920.}
		\label{fig:sa12}
	\end{subfigure}
	\hfill
	\begin{subfigure}[b]{0.55\textwidth}
		\centering
		\includegraphics[width=\textwidth]{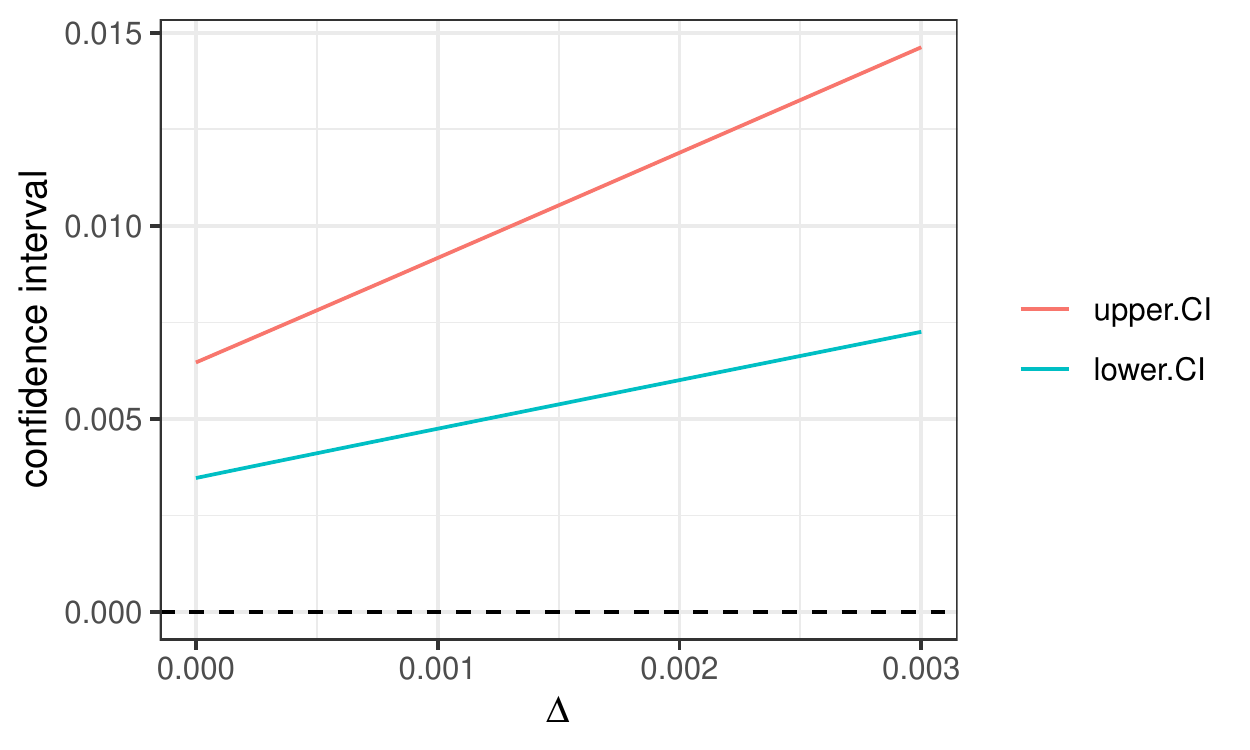}
		\caption{Birth cohorts: 1921-1930.}
		\label{fig:sa23}
	\end{subfigure}
	\hfill
	\begin{subfigure}[b]{0.55\textwidth}
		\centering
		\includegraphics[width=\textwidth]{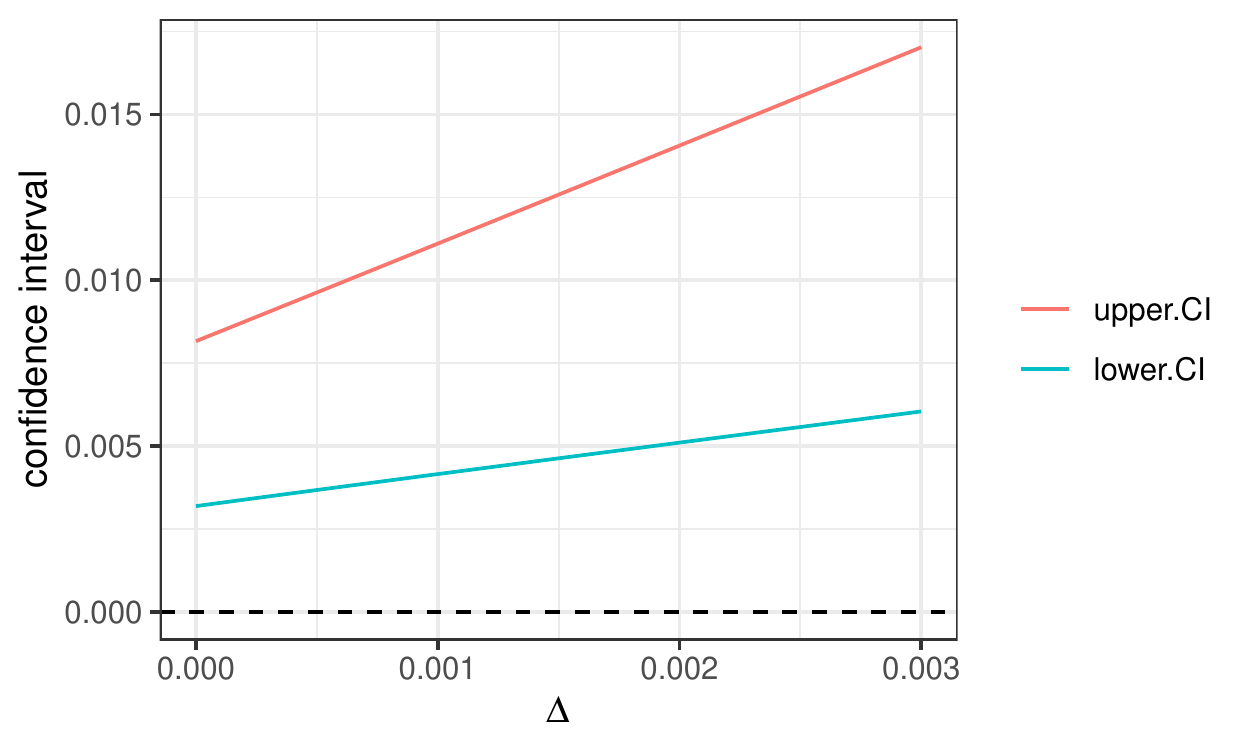}
		\caption{Birth cohorts: 1931-1940.}
		\label{fig:sa34}
	\end{subfigure}
	\caption{Confidence intervals for $ \beta^* $ when $ \Delta\in [0, 0.3\%] $. The confidence intervals do not cover zero, which means that the observed treatment effect cannot be explained away by $ \Delta\in [0, 0.3\%] $.}
	\label{fig:sa}
\end{figure}

Because the concern is that the effect of smoking on lung cancer increases over time, we consider  $ \gamma_L=0 $ and $ \gamma_U=0.3\% $, i.e., we consider every value of 
 $ \Delta\in [0, 0.3\%] $. The constructed confidence intervals for each two consecutive birth cohorts are in Figure \ref{fig:sa}, which indicates that any  $ \Delta\in [0, 0.3\%] $ cannot explain away the treatment effect. In fact, any positive $ \Delta $ cannot explain away the treatment effect. This means that the study conclusion is robust to possible violation of Assumption  \ref{assump: iv for trend}(d). 

\end{document}